# Fast Antibiotic resistance-Based gene editing of mammalian cells with CRISPR-Cas9 (FAB-CRISPR)

**Petia Adarska[1,7], Eleanor Fox[1,7,2], Joshua Heyza[3,4,7], Carlo Barnaba[3,5,7], Jens Schmidt[3,6*] and Francesca Bottanelli[1*]**

[1]Freie Universität Berlin, Institute of Chemistry and Biochemistry, Thielallee 63, 14195 Berlin, Germany

[2]Current Affilitation: Cambridge Institute for Medical Research, University of Cambridge, Cambridge, United Kingdom

[3]Institute for Quantitative Health Science and Engineering, Michigan State University, East Lansing, MI, USA

[4]Current Affilitation: Department of Pharmacology, Wayne State University School of Medicine, Detroit, MI, USA

[5]Current Affilitation: Department of Pharmaceutical Chemistry, University of Kansas, Lawrence, KS, USA

[6]Department of Obstetrics, Gynecology, and Reproductive Biology, Michigan State University, East Lansing, MI, USA

[7]These authors contributed equally

[*]Correspondence: francesca.bottanelli@fu-berlin.de, schmi706@msu.edu

## Summary

Protein tagging with CRISPR-Cas9 enables the investigation of protein function in its native environment but is limited by low homology-directed repair (HDR) efficiency causing low knock-in rates. We present a detailed pipeline using HDR donor plasmids containing antibiotic resistance cassettes for rapid selection of gene-edited cells. Our protocol streamlines N- or C-terminal tagging in human cells, enabling HDR donor plasmid preparation in a single cloning step.

## Graphical abstract

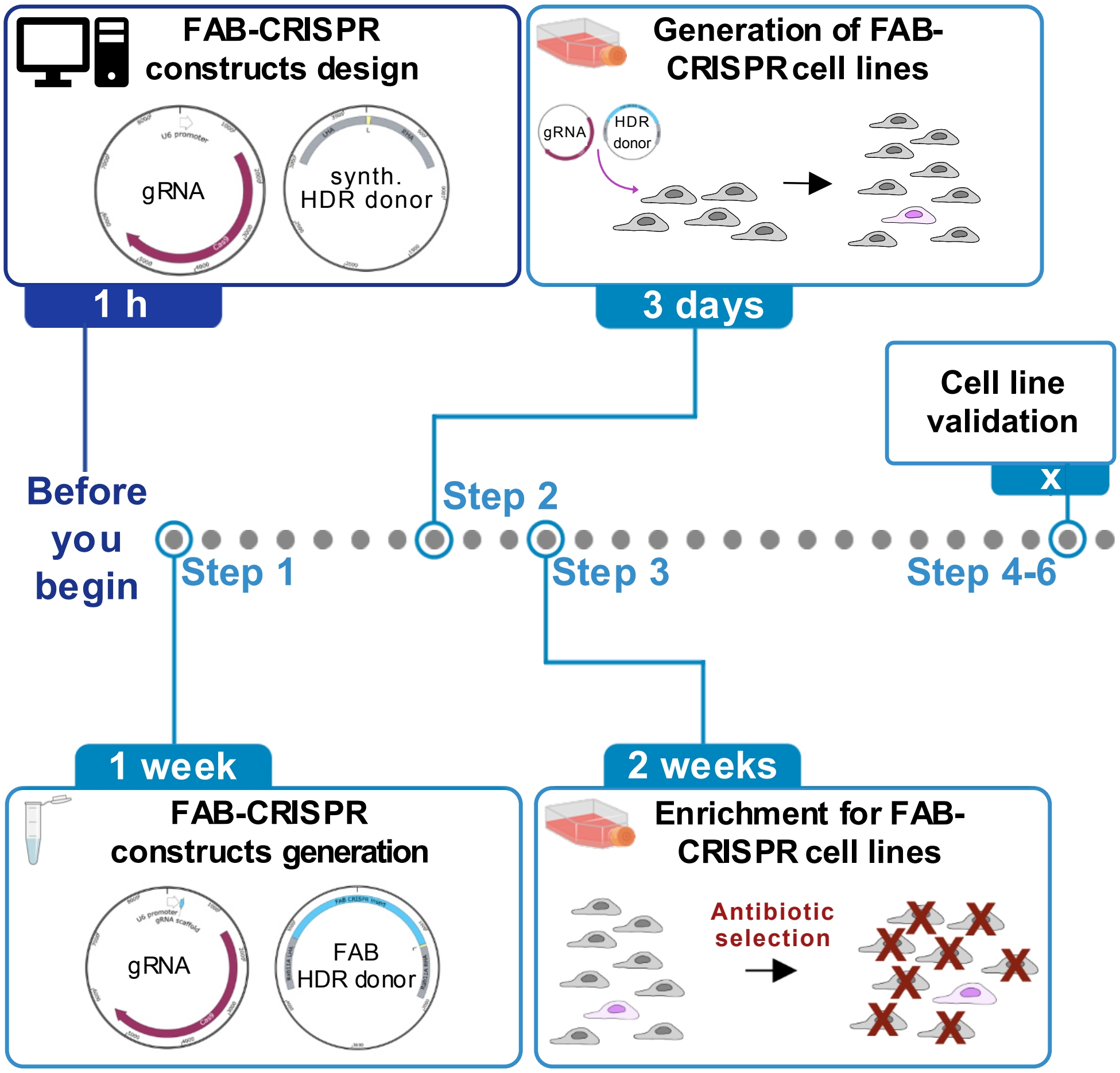

## Before you begin

The site-specific introduction of a DNA double-strand break (DSB) by CRISPR-Cas9 allows for targeted genomic alterations that enable protein knock-out (KO) and knock-in (KI)(Cong et al., 2013; Jinek et al., 2012; Mali et al., 2013). A major hurdle of gene editing is the low efficiency of homology-directed repair (HDR) which results in KI efficiency in the low percentage range. Here, we present a step-by-step protocol for endogenous protein tagging in human cells, which overcomes low HDR efficiency through an integrated antibiotic resistance cassette. Our workflow guides users from reagent design to cell line validation. The provided FAB-CRISPR HDR donor plasmids contain pre-designed inserts for C- and N-terminal tagging, allowing for simple copying and pasting into your choice HDR donor plasmid in a single cloning step. The FAB-CRISPR plasmids contain a variety of Tag and antibiotic resistance cassette combinations (Figure 1), enabling live-cell microscopy [SNAP (Keppler et al., 2003), Halo (Los et al., 2008)] and the new highly photostable green fluorescent protein monomeric StayGold (Ando et al., 2024) as well as proximity-based proteomics (Branon et al., 2018; Stockhammer et al., 2024b) (Turbo-ID). Additionally, SNAP and Halo Tags have been used to acutely modulate protein function using proteolysis targeting chimera (PROTAC) substrates (Buckley et al., 2015; Pol et al., 2024; Tovell et al., 2019).

Our method has been applied to introduce many other protein Tags (*e.g.*, mEOS3.2, enhanced GFP (EGFP), APEX2, and small epitope Tags) into multiple mammalian cell lines, including U2OS, RPE-1, Jurkat T-cells, T84, HAP1, MCF7, MCF10A, and Caco2. The pipeline is highly adaptable for introducing any exogenous DNA fragment into transfection-compatible cell genomes. Genes involved in a wide range of cellular processes, such as membrane trafficking (Bottanelli et al., 2017; Pol et al., 2024; Stockhammer et al., 2024a; Stockhammer et al., 2024b; Wong-Dilworth et al., 2023), autophagy (Barnaba et al., 2024; Broadbent et al., 2023; Broadbent et al., 2025), DNA repair (Brambati et al., 2023; Heyza et al., 2025; Heyza et al., 2023b; Mashayekhi et al., 2024; Mikhova et al., 2024), and telomere maintenance (Janovic et al., 2024; Klump et al., 2023; Schmidt et al., 2016) have also been successfully targeted using this method.

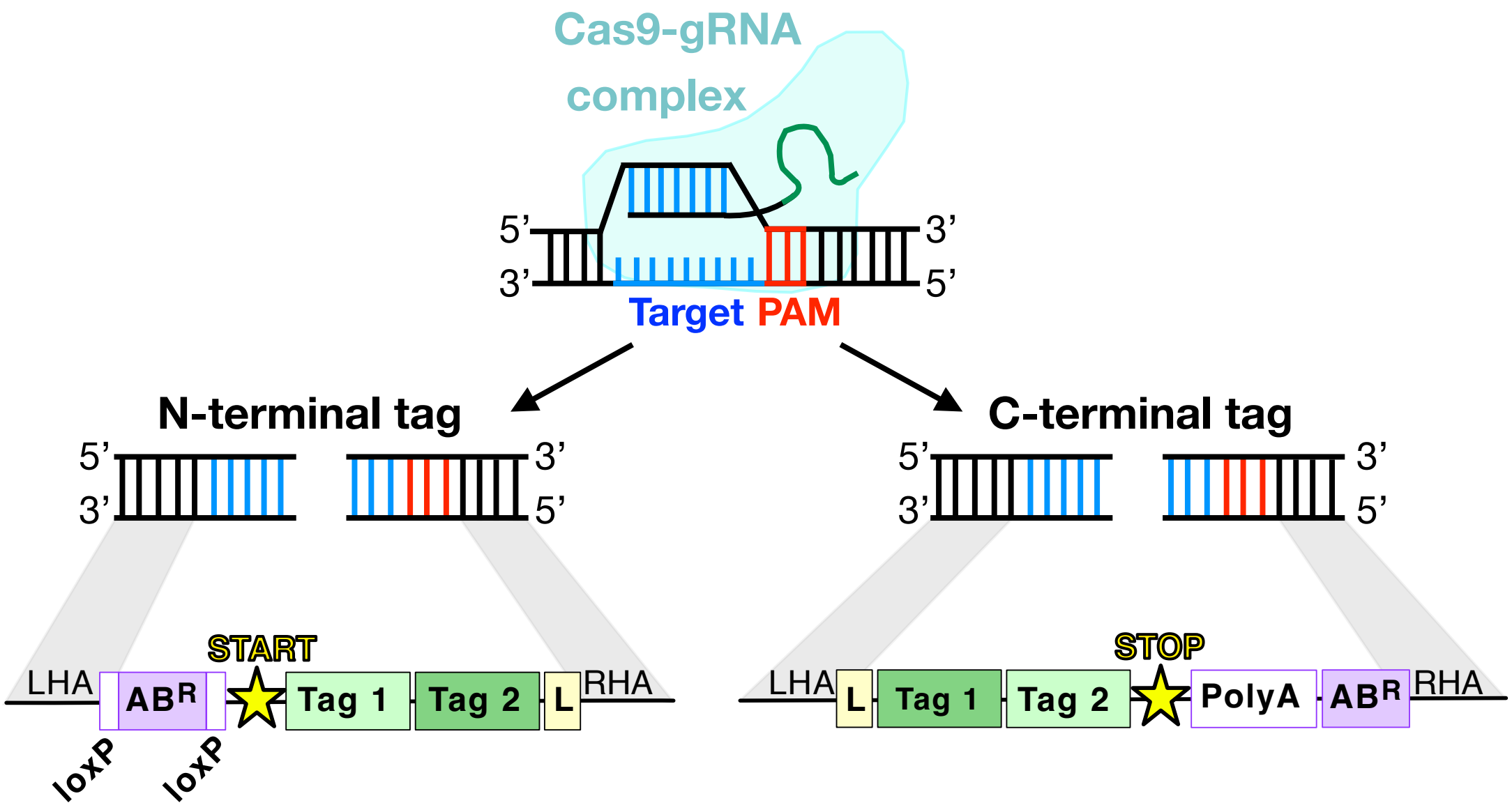

| Antibiotic resistance for CRISPR KI enrichment | Epitope tag (WB, IF) | Tag | Application |
|---|---|---|---|
| G418 | V5 | Turbo-ID | Proximity Biotinylation |
| | ALFA | Halo | Fixed and live-cell imaging (e.g.: confocal, STED, TIRF), Induced degradation (PROTAC) |
| Puromycin | V5 | SNAP | |
| Hygromycin | (HA) | mStayGold | Live-cell imaging (e.g.: confocal, TIRF) |

**Figure 1: Overview of the available Tag-antibiotic resistance cassette (FAB-CRISPR) encoding HDR donor plasmids described in this paper.** We present various FAB-CRISPR cassettes that can be used to generate KIs for various downstream applications including proximity-based proteomics, live-cell microscopy [*e.g.,* stimulated emission depletion (STED) and total internal reflection fluorescence (TIRF)] and induced degradation using proteolysis targeting chimera (PROTAC). Monomeric StayGold (mStayGold), linker (L). Not drawn to scale.

Two components are required to generate KI cell lines. First, a 20 nucleotide long guide RNA (gRNA) is required to target the Cas9 nuclease to the genomic region of interest, leading to the introduction of a DSB, with cleavages on each DNA strand directly opposite each other. Second, a homology-directed repair (HDR) DNA template is necessary to promote homologous recombination of the targeted alleles. The HDR donor plasmid harbors the sequence to be knocked-in, flanked by homology arms (HAs) with homology to the genomic sequences upstream and downstream of the DSB locus. This ultimately leads to error-free, site-specific insertion of the sequence of interest.
In this first section, we will guide you through the process of designing a gRNA and a homology-directed repair (HDR) donor plasmid for the generation of KI cell lines using the FAB-CRISPR cassettes.

## gRNA design for tagging the N- or C-terminus of a protein of interest

**Timing: 30 min.**

Specific considerations must be made to achieve multiple objectives when selecting a suitable gRNA sequence for KIs. These objectives include ensuring accurate insertion of the donor sequence, minimizing the risk of generating indels or point mutations at the insertion site, and preventing cleavage of the HDR donor plasmid to avoid compromising KI efficiency. Additionally, it is crucial to safeguard the expression and biological function of the protein of interest from unintended alterations. Therefore, before starting you should decide at which terminus your protein of interest should be tagged to retain functionality, if this information is available.

1. Go to www.benchling.com and create an account.

2. Import the genomic sequence of interest using Benchling. This is essential to select a gRNA sequence that will target the Cas9 to induce a DSB in an appropriate site and to design the homology-directed repair (HDR) donor plasmid for editing.

3. Create a new project by clicking on the "+" icon (indicated by the red box) (Figure 2).
4. To import the genomic sequence of interest, click on the "+" icon next to your newly created

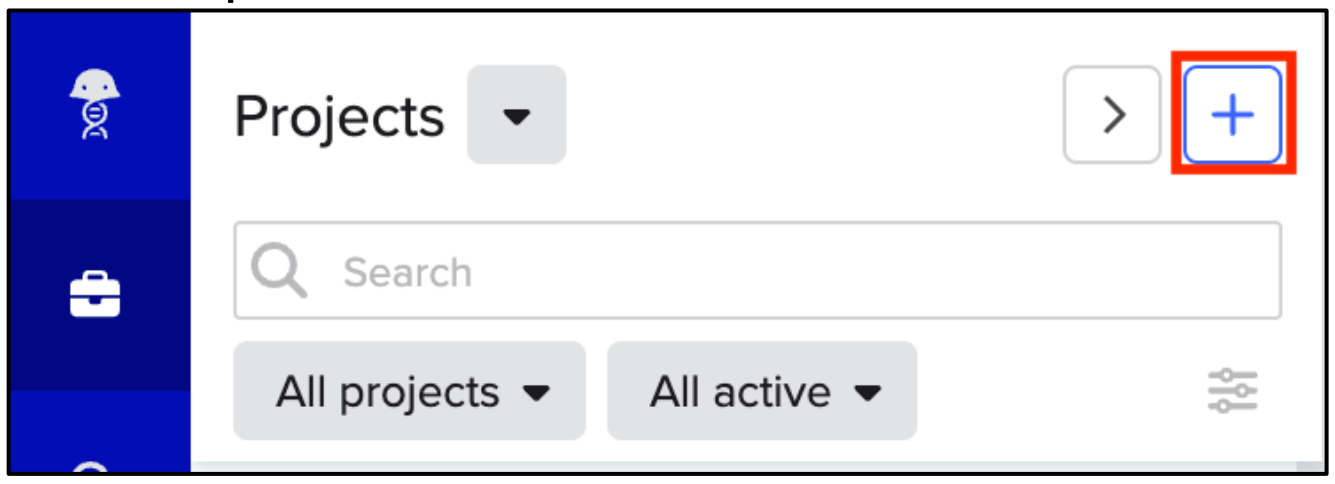

**Figure 2: How to create a new Project in Benchling.**

project and select "DNA/RNA sequence", followed by "import DNA/RNA sequences." Then select "import from external database" and search for your gene of interest in the sequence field, as shown in the example searches (Figure 3). Enter "human" in the genome search field to select the GRCh38 human annotated genome. Click on the Search button.

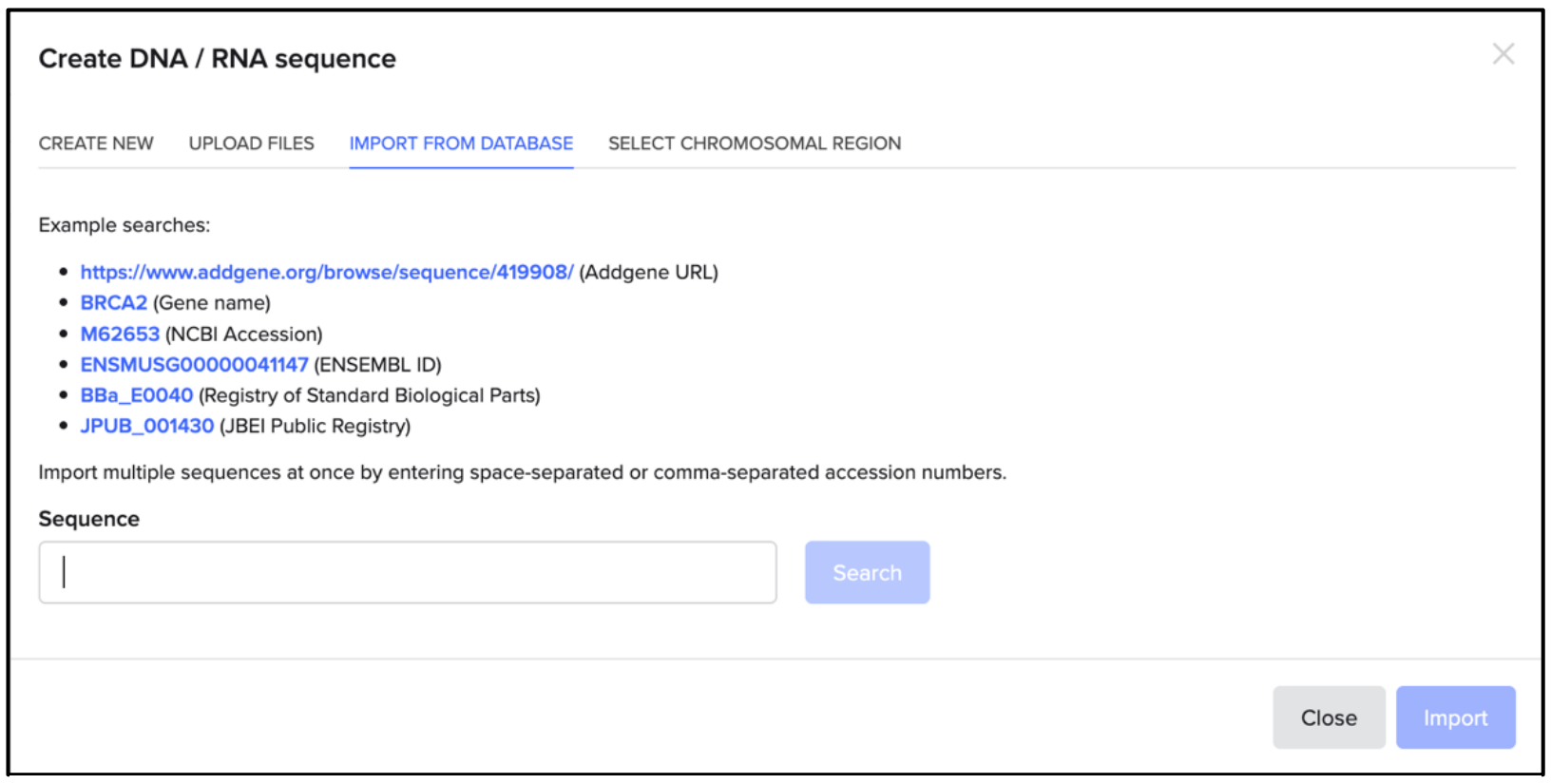

**Figure 3: How to import the desired genomic sequence.**

5. In the next window (Figure 4), you will be able to select which transcript to use for further visualization on Benchling (when multiple transcripts and splicing isoforms are available) and you will get information on the location of your locus of interest (where exactly the sequence of interest is on a specific chromosome). The default transcript is typically the canonical sequence. Import the sequence with an extra 1000 bp upstream and downstream of the locus. This will be important later for the design of the HDR donor plasmid.

**Note:** To obtain information about possible splicing isoforms and transcripts search for your protein of interest on www.uniprot.org.

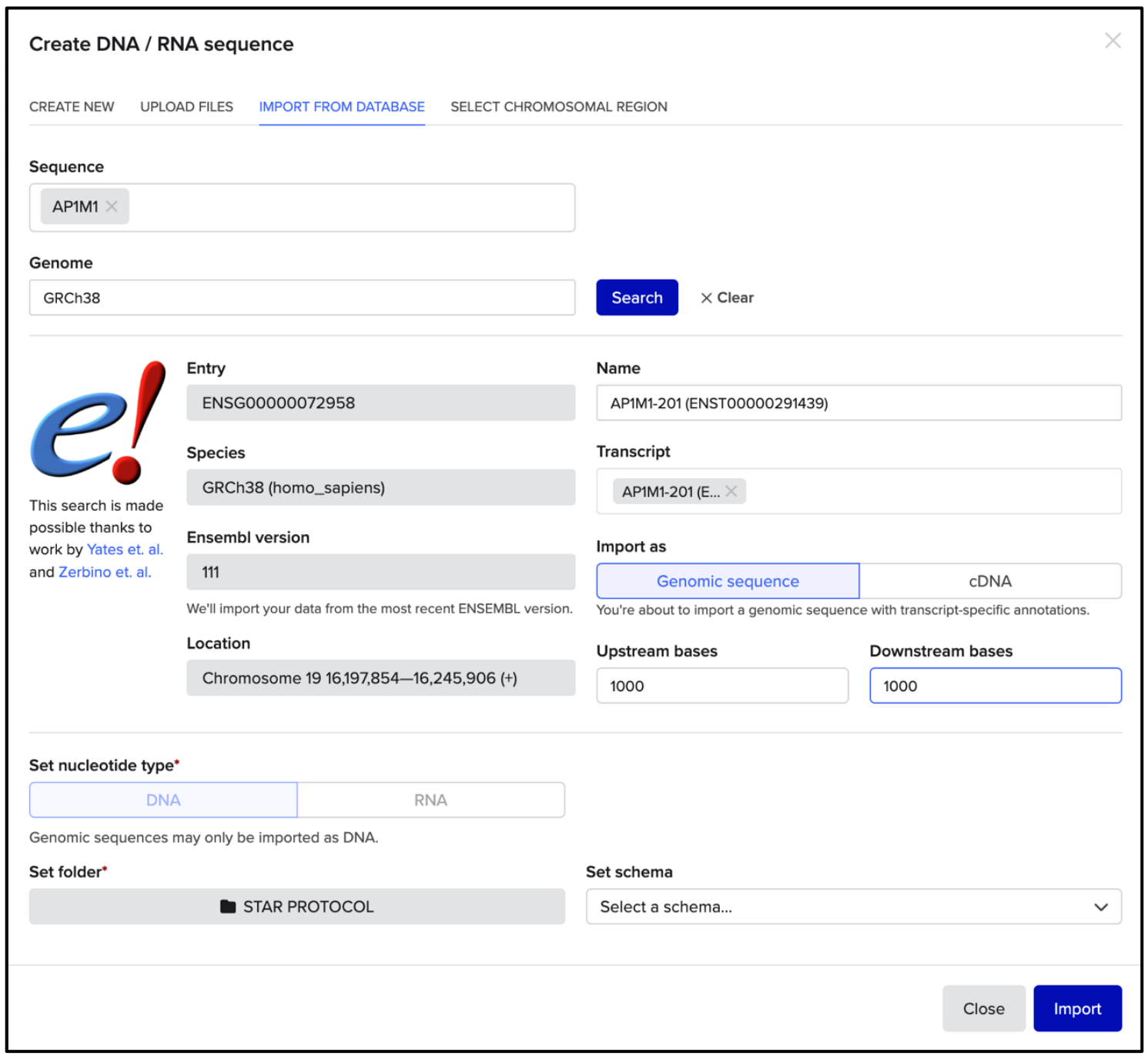

**Figure 4: How to visualize the genomic region of interest for designing the HDR donor plasmid and gRNA.** The AP1M1 gene is used as an example.

6. Identify the START codon (if tagging at the N-terminus) or the STOP codon (if tagging the C-terminus) of the gene of interest (Figure 5). By clicking on the arrows in the linear map (right panel, Figure 5) you will be able to visualize specific exons. In the example shown, we have selected the last exon of the AP1M1 gene locus encoding for AP1μA, a protein involved in intracellular trafficking. The START or the STOP codon will be used to guide the design of the homology arms (HAs) for the HDR donor plasmid as the Tag will be inserted into the genomic locus upstream of the START codon (N-terminal tagging) or downstream of the last aminoacidic-encoding codon (C-terminal tagging). Select the site where you want to insert your Tag (at the N- or C-terminus) and highlight a sequence of +/- 50 bp around the insertion site for designing the gRNA. You can do this by simply selecting the sequence in the sequence map window (left panel, Figure 5) in Benchling. In the example, we have selected a 100 bp sequence around the STOP codon of AP1M1.

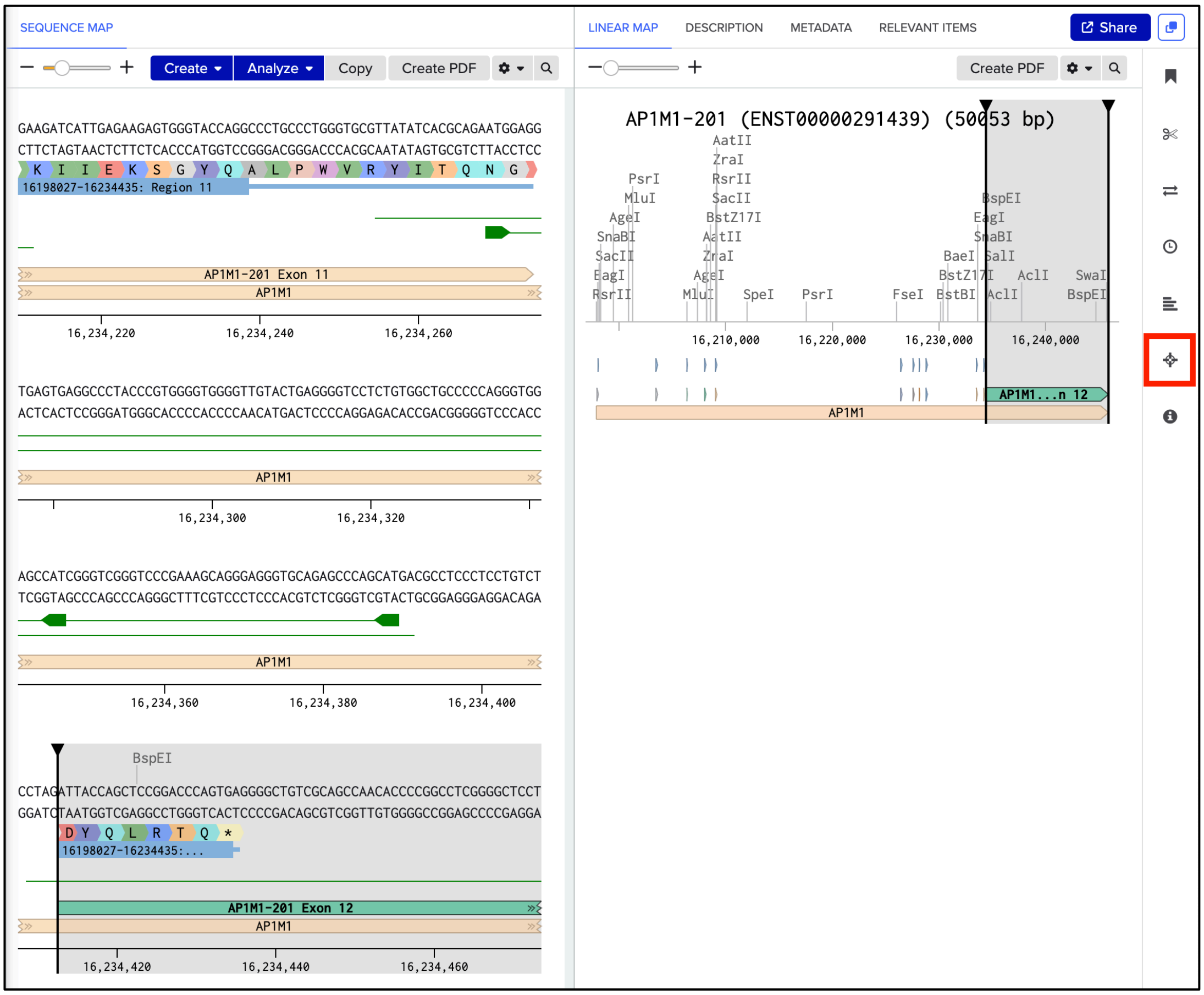

**Figure 5: How to design a gRNA with Benchling.** Click the star shaped icon (highlighted by a red box) for designing gRNAs targeting the selected genomic region. Here the C-terminal region of AP1M1 is visualized.

7. Click on the star-shaped icon on the very right of the page (CRISPR button, highlighted by a red square) and select “design and analyze guides”. Select guide parameters as in the example shown below when using wild-type Cas9 from *Streptococcus pyogenes* Cas9 (SpCas9) (Figure 6). All the visualized guides will contain a PAM sequence (NGG for SpCas9) at the 3’ end as a PAM sequence is necessary for the recognition of the target sequence by the Cas9 complex.

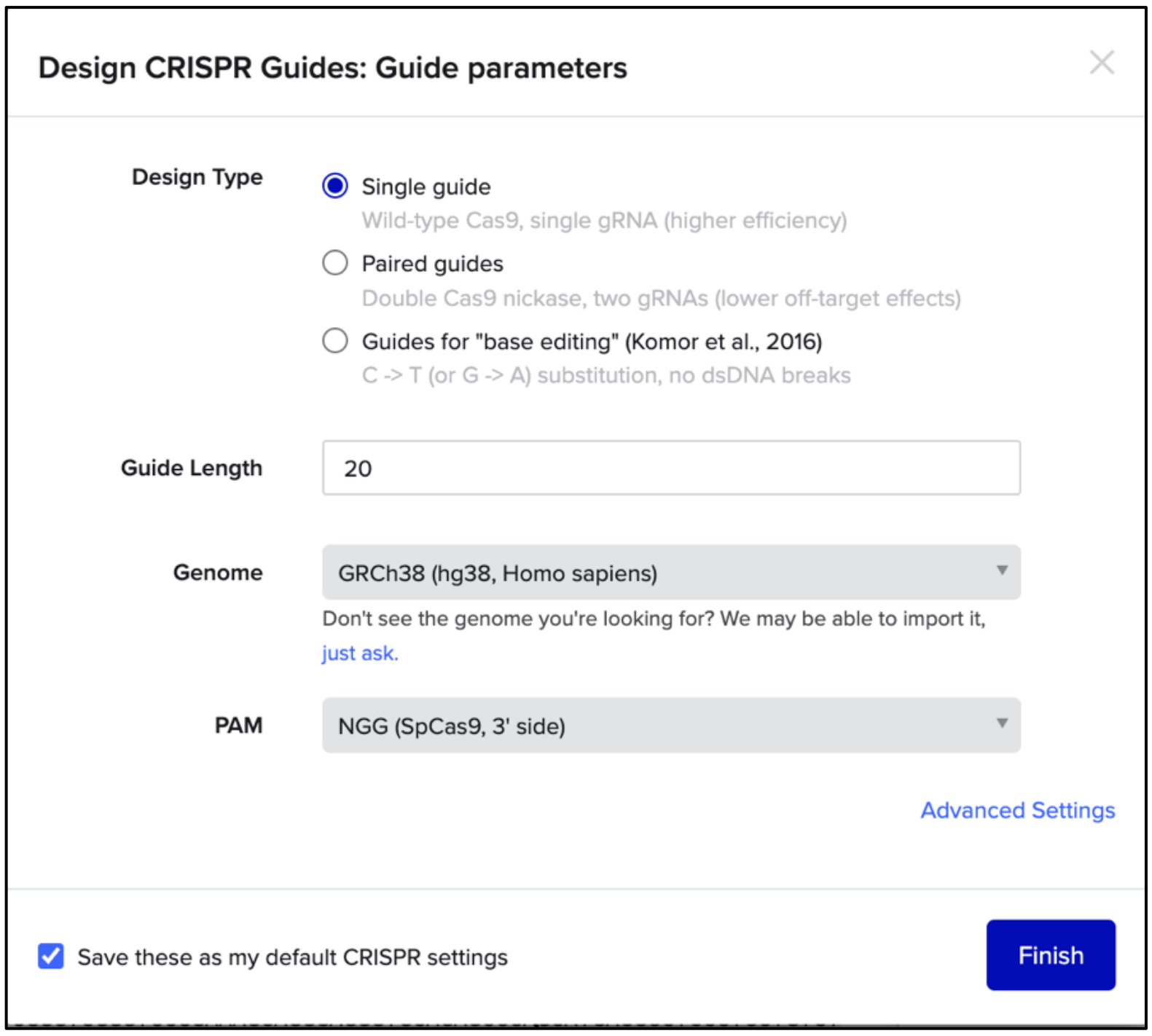

**Figure 6: Parameters for designing a gRNA cutting in the selected genomic region when using *Streptococcus pyogenes* Cas9 (SpCas9).**

8. Click on “Finish” to confirm the selected parameters. This will generate a list of putative gRNAs target sequences in your selected region on either the sense (+) or antisense (-) strands.

**Note:** When working with guide RNAs on the antisense (-) strand remember to design your oligonucleotides from 5’ to 3’.

9. A list of available gRNA target sequences is shown on the right (Figure 7). When clicking on a specific guide sequence, it will be visualized in the sequence map window on the left. We recommend choosing two guides. For choosing an optimal guide consider closeness to the insertion site, the highest on-target score (Doench et al., 2016) and no off-targets (Hsu et al., 2013).

**Critical: Check that no other genes are targeted by the selected guide. The specificity of the selected guide is indicated by an off-target score (a high off-target score is better). Click on the off-target score of the specific guide you are interested in for visualizing possible off-target sequences and the probability of off-target effects.**

**Note:** For N-terminal tagging, we recommend choosing a gRNA that cuts after the START codon. The efficiency of Cas9 cutting and thus non-homologous end-joining (NHEJ) that leads to allele KO is very high, whereas HDR is a rare event. If cutting by the Cas9 occurs after the START codon, cells with one KI allele, where HDR has occurred, and one KO allele, where NHEJ has occurred, will be created. Despite this heterozygosity at the gene level, these cells may appear homozygous at the protein level (to be confirmed via Western Blot). For C-terminal tagging, it is important to choose a guide that will cut after the STOP codon. If cutting by the Cas9 occurs before the STOP codon, heterozygous cells will bear one edited allele and, with high probability, a second allele with mutations and/or frameshifting indels at the very 3' of the open reading frame (ORF). This is undesirable when wanting to work with a mixed population of cells.

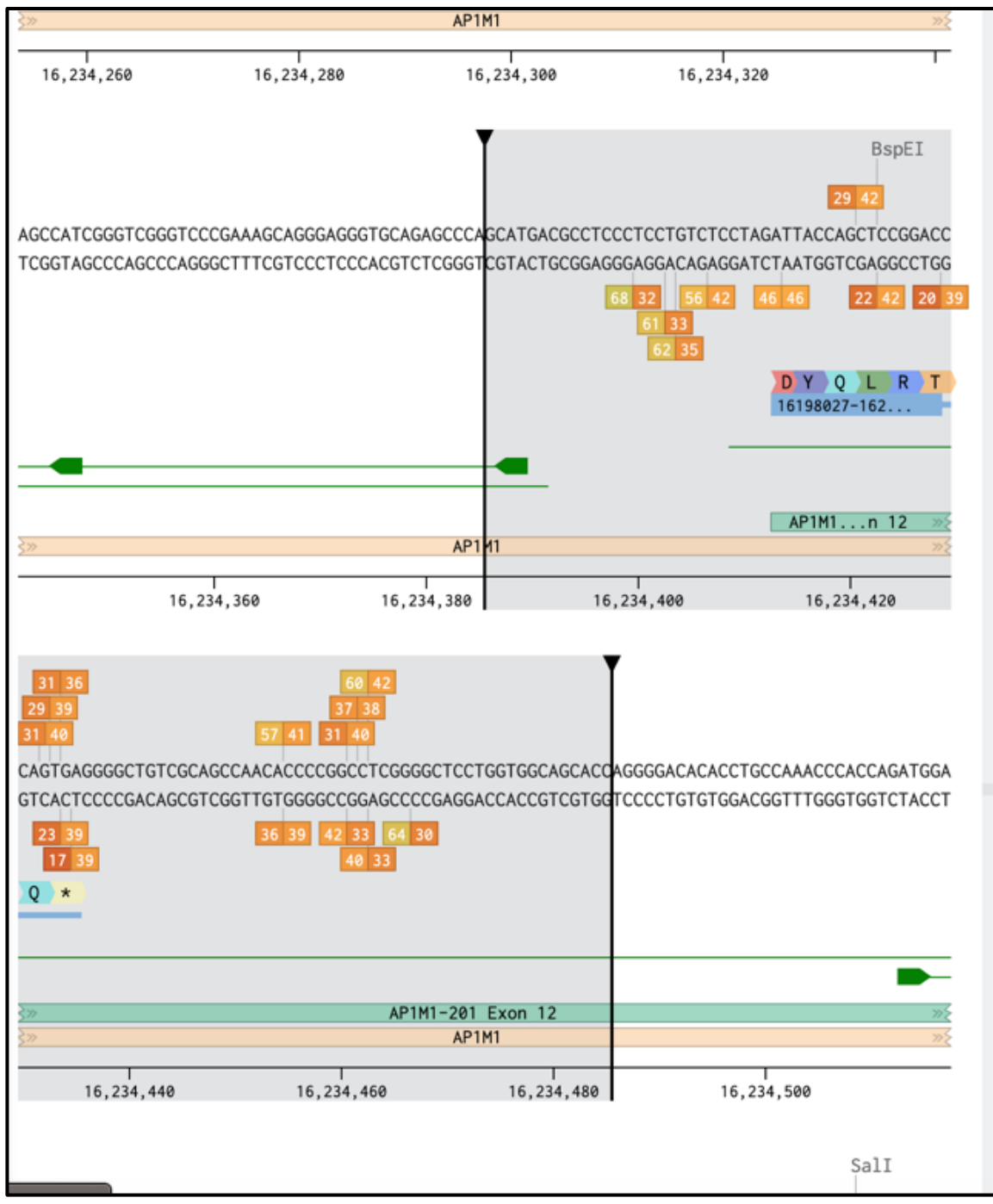

| | Cut Position | Strand | Guide Sequence | PAM | On-Target Score | Off-Target Score |
|---|---|---|---|---|---|---|
| ☐ | 16234400 | - | TAATCTAGGAGACAGGAGGG | AGG | 68.8* | 32.3 |
| ☐ | 16234403 | - | TGGTAATCTAGGAGACAGGA | GGG | 61.9* | 33.8 |
| ☐ | 16234404 | - | CTGGTAATCTAGGAGACAGG | AGG | 62.1* | 35.8 |
| ☐ | 16234407 | - | GAGCTGGTAATCTAGGAGAC | AGG | 56.4* | 42.5 |
| ☐ | 16234414 | - | GGGTCCGGAGCTGGTAATCT | AGG | 47.0* | 46.1 |
| ☐ | 16234421 | + | GTCTCCTAGATTACCAGCTC | CGG | 29.6 | 42.9 |
| ☐ | 16234423 | - | CCCCTCACTGGGTCCGGAGC | TGG | 22.2 | 43.0 |
| ☐ | 16234429 | - | CGACAGCCCCTCACTGGGTC | CGG | 20.9 | 39.4 |
| ☐ | 16234432 | + | TACCAGCTCCGGACCCAGTG | AGG | 31.5 | 40.9 |
| ☐ | 16234433 | + | ACCAGCTCCGGACCCAGTGA | GGG | 30.0 | 39.6 |
| ☐ | 16234434 | + | CCAGCTCCGGACCCAGTGAG | GGG | 31.0 | 36.5 |
| ☐ | 16234434 | - | GGCTGCGACAGCCCCTCACT | GGG | 23.8 | 39.7 |
| ☐ | 16234435 | - | TGGCTGCGACAGCCCCTCAC | TGG | 17.8 | 39.2 |
| ☐ | 16234455 | + | GGCTGTCGCAGCCAACACCC | CGG | 57.5* | 41.8 |
| ☐ | 16234455 | - | GAGCCCCGAGGCCGGGGTGT | TGG | 36.9* | 39.7 |
| ☐ | 16234461 | + | CGCAGCCAACACCCCGGCCT | CGG | 31.3* | 40.8 |
| ☐ | 16234461 | - | CACCAGGAGCCCCGAGGCCG | GGG | 42.6* | 33.5 |
| ☐ | 16234462 | + | GCAGCCAACACCCCGGCCTC | GGG | 37.1* | 38.4 |
| ☐ | 16234462 | - | CCACCAGGAGCCCCGAGGCC | GGG | 39.9* | 24.7 |

**Figure 7: How to select a gRNA.** List of possible gRNAs cutting around the C-terminus of AP1M1.

10. If possible, choose a gRNA sequence that spans the STOP (or START) codon, meaning the complete gRNA sequence would not be included in the HDR donor plasmid, protecting it from enzymatic cleavage by the Cas9. This will also simplify HDR donor plasmid design (see next major Step), as the introduction of a silent mutation in the region of homology to remove the PAM site is not required (Figure 8).

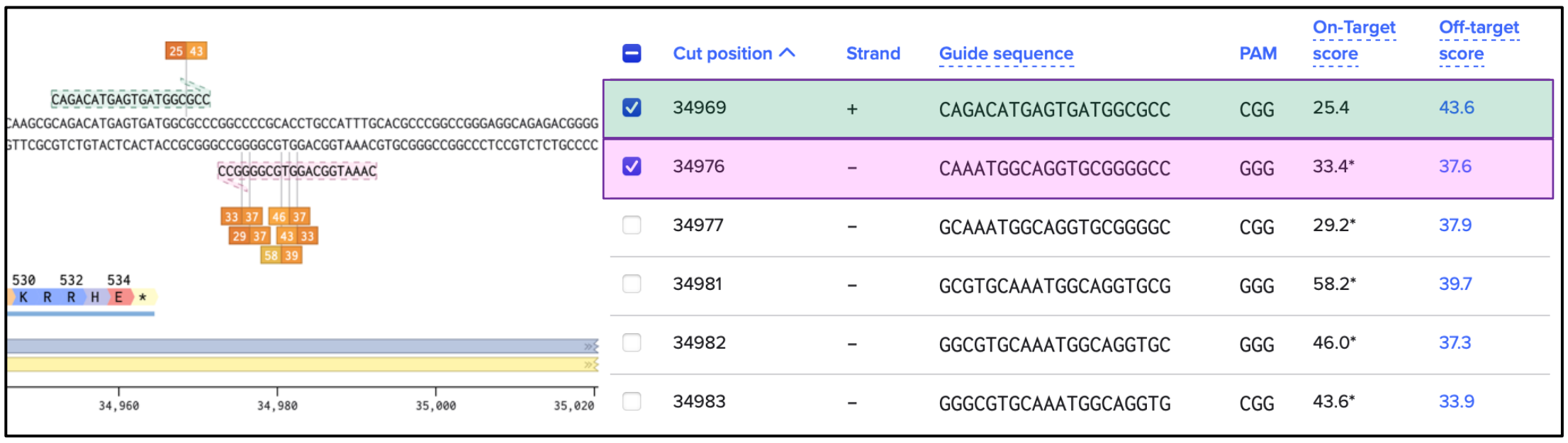

**Figure 8: Example showing possible gRNA targeting the AP1M1 locus.** The green gRNA spans over the STOP codon (on the + strand) so that no silent mutation of the PAM site in the HDR donor plasmid is needed. The magenta gRNA (on the - strand) recognizes a target sequence after the STOP codon, requiring the introduction of silent mutations in the HDR donor plasmids to prevent cutting by the Cas9.

11. Once a gRNA sequence is selected, order two complementary oligonucleotides (oligos), which must include the 20 bases of the gRNA and the additional overhangs for cloning into a pX330 plasmid for simultaneous expression of SpCas9 and the gRNA (Figure 9) (Cong et al., 2013; Ran et al., 2013).

**Guide annealing (to insert into pX330 plasmid)**

5' CACCGNNNNNNNNNNNNNNNNNNNN 3'
3' CNNNNNNNNNNNNNNNNNNNNCAAA 5'

**Figure 9: How to design oligos for gRNA cloning into the pX330 plasmid.** For cloning into the gRNA scaffold of pX330 the designed oligonucleotides need overhangs (indicated in red) that allow ligation of the annealed oligos into the pX330 plasmid digested with BpiI.

**Note:** Ordering the correct oligos is critical to ensure proper annealing.
Oligonucleotide #1: CACCGNNNNNNNNNNNNNNNNNNNN
Oligonucleotide #2: AAACNNNNNNNNNNNNNNNNNNNNC

## Designing a homology-directed repair donor (HDR) plasmid for N- and C-terminal tagging

**Timing: 30 min.**

For tagging a protein of interest, we have implemented a single-step (C-terminal tagging) or two-step (N-terminal tagging) genome editing strategy. The first step allows for antibiotic selection of positively edited cells that have integrated a sequence present on an HDR donor plasmid (a sequence encoding for a Tag and an antibiotic selection cassette). For C-terminal protein tagging, the antibiotic resistance cassette is placed downstream of the Tag and an exogenous polyadenylation (polyA) sequence. For N-terminal fusions, the selection marker is upstream of the Tag, or within the Tag coding sequence, and is flanked by LoxP sites allowing for removal of the antibiotic resistance gene by EGFP-Cre mediated recombination (Xi et al., 2015).

For each target gene, a HDR donor plasmid consisting of a left homology arm (LHA, ~500 bp), two unique restriction sites for cloning of the Tag and the antibiotic resistance cassette and a right homology arm (RHA, ~500 bp) are synthesized using your favorite DNA synthesis service. With a single cloning step, you will then be able to copy and paste a Tag-polyA-$AB^R$ (antibiotic resistance cassette; for C-terminal fusions) or a LoxP-$AB^R$-LoxP-Tag (for N-terminal fusions) from the provided FAB-CRISPR plasmids collection (described later in Major Step 1b and 1c, for an overview of the available FAB-CRISPR Tags see Figure 1). Alternatively, Gibson assembly or standard genomic polymerase chain reaction (PCR) can be used to clone the HAs into your plasmid of choice (Figure 10). Plasmids can be assembled *in silico* in a molecular biology software such as SnapGene to generate a target sequence map.

12. Use Benchling to select 500 bp sequences downstream/upstream of the START codon (N-terminal tagging) or downstream/upstream from the STOP codon (C-terminal tagging).

13. Make sure to include two restriction sites in between the two HAs to be able to copy and paste Tags and resistance markers from the suggested FAB-CRISPR plasmids (as described later in 1b and 1c). For N-terminal tagging, design the HDR donor plasmid as LHA-NheI-NNNNN-BamHI-L-RHA. For C-terminal tagging design the HDR donor plasmid as LHA-L-BamHI-NNNNN-EcoRI-RHA, where NNNNN are random bases to facilitate restriction digests and L is a glycine-serine linker of choice. The HDR donor plasmid can be synthesized from any nucleic acid supplier as a DNA sequence in a high copy plasmid with the preferred bacterial antibiotic resistance.

**Critical: Before proceeding with the synthesis, verify that the *in silico* assembled HDR donor plasmid contains a START codon immediately preceding the Tag for N-terminal fusions. For C-terminal fusions, verify the presence of a STOP codon at the end of the Tag fragment.**

**Critical: When synthesizing your HDR plasmid, make sure to order a high copy number plasmid, such as pUC plasmids.**

**Critical: If one of the homology arms contains the gRNA sequence, the Cas9 will cleave the HDR donor plasmid (see Figure 8). To overcome this issue, a mutation should be introduced to disrupt the PAM sequence on the HDR donor plasmid. When mutating the PAM sequence is not possible, introduce 2-3 single point mutations in the upstream 20 base sequence recognized by the Cas9. In the case of N-terminal tagging, it is important to introduce a silent mutation that will not alter the protein product. (If no silent mutations can be introduced, choose another gRNA.)**

**Critical: If the suggested restriction enzymes cut within your designed HDR donor plasmid, mutagenize the recognition sequences by inserting silent point mutations (within the coding sequence). We have successfully introduced mutations in introns without altering the levels of downstream protein products. Alternatively, use different restriction sites and amplify the Tag and antibiotic resistance cassettes via PCR.**

**Note:** In both the N- and C-terminal tagging strategies, a TEV protease cleavage site can be added beside the flexible linker, which allows the removal of the Tag from the protein to provide flexibility for downstream experiments (Heyza et al., 2023a; Schmidt et al., 2016).

**Alternative:** In some instances (*e.g.,* GC-rich sequences) DNA synthesis may not be possible. In this case, HAs can be obtained as double-stranded DNA fragments and cloned via Gibson assembly or amplified directly from genomic DNA (see the alternative protocol in Annex 1). We have successfully used HA, which are of lengths 250-450 bp downstream/upstream of the START codon (N-terminal tagging) or downstream/upstream from the STOP codon (C-terminal tagging) (Broadbent et al., 2023; Heyza et al., 2023a). The HAs will be cloned into the HDR plasmid using Gibson assembly; thus, they will need to contain a sequence of 20-25 bp on both sides that overlap with the HDR plasmid backbone and the Tag sequence (Figure 10).

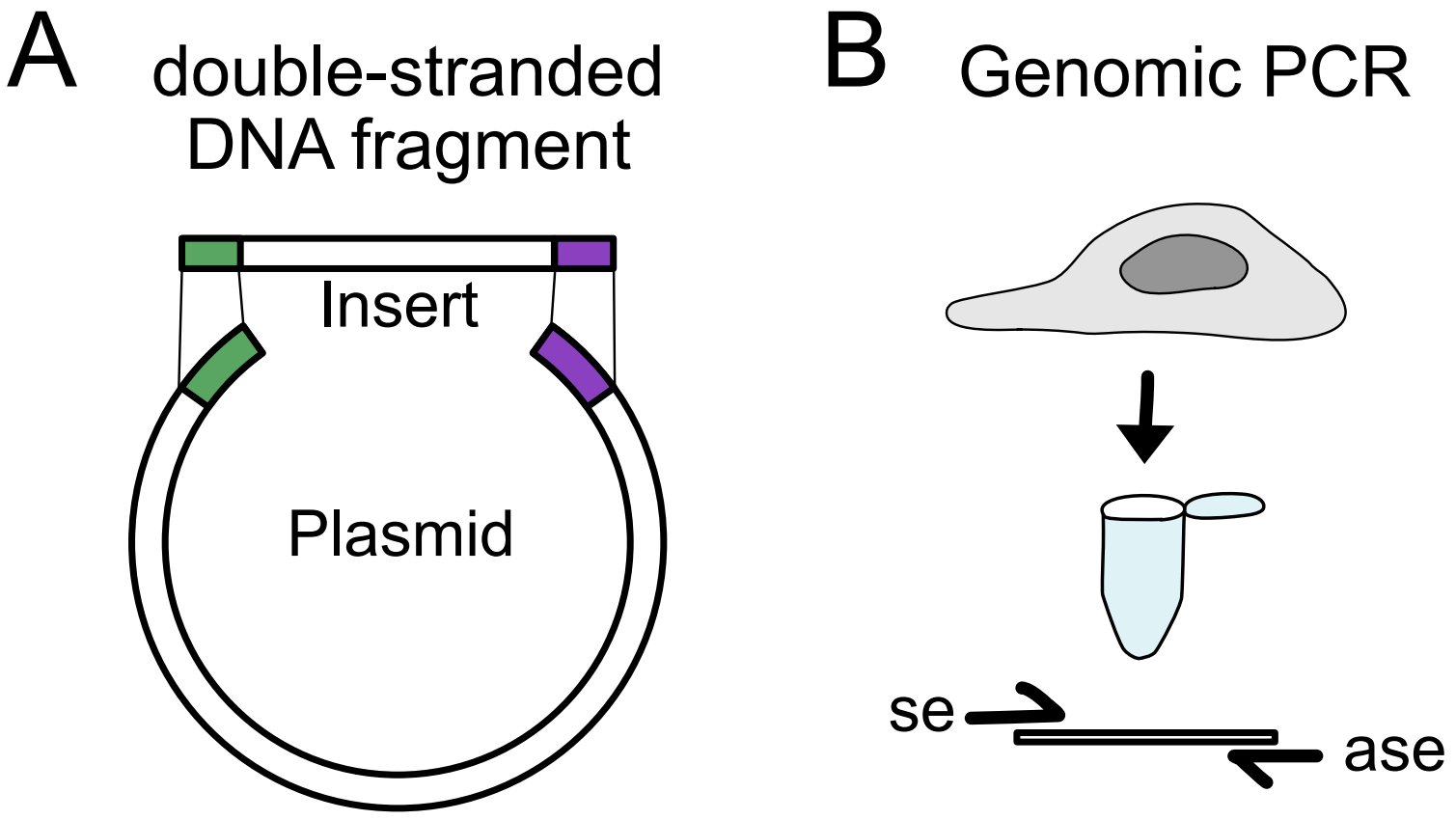

**Figure 10: Alternative methods for obtaining homology arms.** (**A**) HAs can be synthesized as a double-stranded DNA fragment containing overlapping sequences with the backbone plasmid of choice (indicated in green and violet) to allow cloning via Gibson assembly. (**B**) Alternatively, HAs can be amplified from genomic DNA via PCR for restriction-digest-based cloning. Polymerase chain reaction (PCR), se (sense), ase (anti-sense).

## List of abbreviations

$AB^R$ – Antibiotic resistance cassette
Ase - Anti-sense
BME - Beta-mercaptoethanol
BSA - Bovines serum albumin
CIP - Calf intestinal alkaline phosphatase
CRISPR - Clustered regularly interspaced short palindromic repeats
DMEM - Dulbecco's modified eagle's medium
DSB - Double-strand break
E. coli - *Escherichia coli*
EDTA - Ethylenediaminetetraacetic acid
FACS - Fluorescence activated cell sorting
FBS - Fetal bovine serum
G418 - Geneticin
gRNA - Guide RNA
HA - Homology arm
HDR - Homology-directed repair
KI - Knock-in
KO - Knock-out
LAP - Localization and affinity purification
LB - Lysogeny Broth
LHA - Left homology arm
mStayGold - Monomeric StayGold
Oligos - Oligonucleotides
ON - Overnight
PBS - Phosphate-buffered saline
PBST - Phosphate-buffered saline with Tween-20
PCR - Polymerase chain reaction
polyA - Polyadenylation
PROTAC - Proteolysis targeting chimera
RHA - Right homology arm
RS - Restriction site
RT - Room temperature
Se - Sense
STED - Stimulated emission depletion
TIRF - Total internal reflection fluorescence
WT- Wild type

## Key resources table

| REAGENT or RESOURCE | SOURCE | IDENTIFIER |
|---|---|---|
| **Antibodies** | | |
| Anti-ALFA mouse | NanoTag | N1582 |
| Anti-HA mouse | Promega | G921A |
| Anti-V5 rabbit | Abcam | ab84340 |
| Anti-Halo mouse | Promega | G921 |
| Anti-SNAP rabbit | Genscript | A00684 |
| HRP anti-rabbit antibody | Abcam | AB6721 |
| HRP anti-mouse antibody | Abcam | AB6789 |
| **Bacterial and virus strains** | | |
| Top10 *E. coli* competent cells | Life Technologies GmbH | C404003 |
| **Chemicals, peptides, and recombinant proteins** | | |
| FastDigest BpiI enzyme | Thermo Scientific | FD1014 |
| Dulbecco's Modified Eagle's Medium – high glucose (DMEM) | ThermoFisher | 41966029 |
| Fetal bovine serum (FBS) | Corning | 35-079-CV |
| FluoroBrite™ DMEM | Thermo scientific | A1896701 |
| FuGENE(R) HD Transfection Reagent | Promega | E2311 |
| Geneticindisulfate (G418)-solution | Gibco | 11811-098 |
| GlutaMAX | Gibco | 35050061 |
| HEPES PUFFERAN® | Carl Roth | 9105.2 |
| OptiMEM | ThermoFisher | 31985070 |
| Penicillin-Streptomycin | FisherScientific | DE17-602E |
| Phosphate-buffered solution (PBS) | Lonza | 17-156F |
| Bovines serum albumin (BSA) | Carl Roth | 8076.4 |
| 20X Bolt MES SDS Running Buffer | Life Technologies GmbH | B000202 |
| Roti®fair TG-Western | Carl Roth | 1276.1 |
| Puromycin | Life Technologies GmbH | A1113803 |
| Quick calf intestinal alkaline phosphatase (CIP) | New England Biolabs | M0491S |
| Trypsin-EDTA solution | Lonza | 17-161E |
| T4 DNA ligase | New England Biolabs | M0202S |
| Deoxynucleotide (dNTP) | New England Biolabs | N0447S |
| Phusion Polymerase | Thermo scientific | F530S |
| **Critical commercial assays** | | |
| Wizard Plus SV Minipreps DNA Purification System | Promega GmbH | A1460 |
| QIAquick Gel Extraction Kit | Qiagen | 28706 |
| QIAquick PCR Purification Kit | Qiagen | 28106 |
| QIAquick DNA Extraction Kit | Lucigen | QE09050 |
| SuperSignal West Pico Plus | FisherScientific | 34579 |
| **Experimental models: Cell lines** | | |
| HeLa CCL-2 | ATCC | N/A |
| **Oligonucleotides** | | |
| Designed guide RNA/primer sequences | Integrated DNA technologies | N/A |
| **Recombinant DNA** | | |
| pBS598 EF1alpha-EGFPcre | (Le et al., 1999) | Addgene plasmid #11923 |
| pX330-U6-Chimeric_BB-CBh-hSpCas9 | (Cong et al., 2013) | Addgene plasmid #42230 |
| AP1muA single guide RNA | (Stockhammer et al., 2024b) | Addgene plasmid #230030 |
| AP1muA-TurboID-polyA-G418 HDR donor plasmid | | Addgene plasmid #227734 |
| AP1muA-Halo-ALFA-polyA-G418 HDR donor plasmid | (Stockhammer et al., 2024a) | Addgene plasmid #229681 |
| AP1muA-mStayGold-polyA-Hygromycin HDR donor plasmid | | Addgene plasmid #229679 |
| AP1muA-SNAP-V5-polyA-Puro HDR donor plasmid | | Addgene plasmid #229680 |
| Rab11 single guide RNA | | Addgene plasmid #229683 |
| LoxP-G418-LoxP-3xALFA-Halo-Rab11 HDR donor plasmid | | Addgene plasmid #229676 |
| LoxP-Puro-LoxP-3xV5-SNAP-Rab11 HDR donor plasmid | | Addgene plasmid #229677 |
| LoxP-Hygromycin-LoxP-HA-mStayGold-Rab11 HDR donor plasmid | | Addgene plasmid #229678 |
| LoxP-G418-LoxP-TurboID-Rab11 HDR donor plasmid | This paper | Addgene plasmid #230026 |
| **Software and algorithms** | | |
| Benchling | Benchling | www.benchling.com |
| SnapGene Viewer | SnapGene | www.snapgene.com |
| **Other** | | |
| Glass Bottom Dish 35 mm with 14 mm micro-well, #1.5 cover glass | IBL | D35-14-1.5-N |
| Janelia Fluor 646 HaloTag Ligand | Promega | GA1120 |
| SNAP-Cell 647-SiR ligand | New England Biolabs | S9102S |
| Janelia Fluor JFX650 HaloTag ligand | Promega | HT1070 |

# Materials and equipment

# Buffers for cloning

**Annealing buffer**

| Reagent | Final concentration |
| --- | --- |
| Tris, pH 7.5-8.0 | 10 mM |
| NaCl | 50 mM |
| EDTA | 1 mM |

This solution should be made in autoclaved water and can be stored at room temperature (RT) for up to 1 year.

**TE Buffer**

| Reagent | Final concentration |
| --- | --- |
| Tris-HCl pH 6.8 | 10 mM |
| EDTA pH 8 | 1 mM |

This solution should be made in autoclaved water and can be stored at RT for up to 1 year.

# Media for cell culture

**Growth medium**

| Reagent | Final concentration | Amount |
| --- | --- | --- |
| FBS | 10% | 50 mL |
| Penicillin/streptomycin | 1% | 5 mL |
| DMEM | N/A | 445 mL |

This solution needs to be made in sterile conditions and stored at 4 °C.

**Live-cell imaging solution**

| Reagent | Final concentration | Amount |
| --- | --- | --- |
| GlutaMAX (100X) | 1% | 5 mL |
| HEPES | 2% | 10 mL |
| FBS | 10% | 50 mL |
| FluoroBrite DMEM | N/A | 435 mL |

Media needs to be made in sterile conditions and stored at 4 °C.

# Buffers for Western Blot

**2X Laemmli Buffer**

| Reagent | Final concentration |
| --- | --- |
| SDS | 4% |
| Glycerol | 20% |
| Bromophenol blue | 0.02% |
| Tris-HCl pH=6.8 | 120 mM |
| BME | 5% |

This solution can be made as a stock without BME and stored at RT (we recommend adding BME on the day of the sample preparation, once BME is added the solution can be stored at -20 °C).

**Critical: BME should only be opened under a fume hood.**

**PBST**

| Reagent | Final concentration |
|---|---|
| Tween 20 | 0.5% |
| 1X PBS | N/A |

This solution can be stored at RT.

**Blocking buffer**

| Reagent | Final concentration |
|---|---|
| Milk powder | 5% |
| BSA | 1% |
| PBST | N/A |

This solution must be made freshly.

**Alternatives**:

- For simplicity, this protocol is written for editing of HeLa cells, and plasmid transfection was carried out with FuGene. Other cell lines can be successfully edited if efficient transfection can be achieved via any transfection methods. We have successfully edited various cell lines (HAP1, MCF7, MCF10, T84, Caco2, Jurkat, RPE-1, U2OS) using a NEPAgene NEPA21 electroporation machine.
- Substitute any molecular biology reagents successfully used in your laboratory (*e.g.,* polymerase, DNA extraction kits, chemically competent cells, etc.).
- Other antibiotics have successfully been used by our laboratories (Zeocin, Blasticidin). Many cell lines may have been immortalized with plasmids harboring antibiotic resistance cassettes (*e.g.,* RPE-1 are resistant to Hygromycin and Puromycin). Always check which antibiotics work for your cell line of choice.
- Double-KIs can be created simultaneously (transfecting WT cells with two gRNAs and two HDR donor plasmids, each targeting a different locus) or sequentially (transfect an existing single CRISPR KI cell line with a gRNA and HDR donor plasmid, which targets a different locus). To ensure an efficient selection of edited cells, we recommend using a different antibiotic resistance cassette for each HDR donor plasmid (*e.g.*, Puromycin, G418, Hygromycin). Cells can then be treated with optimized concentrations of both antibiotics, which may improve efficiency.

## Step-by-step method details

1a. Cloning of the gRNA in the pX330 plasmid
1b. Generation of HDR donor plasmids for N-terminal tagging
1c. Generation of HDR donor plasmids for C-terminal tagging
2. Transfection of editing reagents into HeLa cells
3a. Selection of edited cells (C-terminal tagging)
3b. Selection of edited cells (N-terminal tagging)
4. Verification of successfully edited cells
5. Further selection of edited cells
6. Functional verification of the endogenously tagged protein

## 1a. Cloning of the gRNA in the pX330 plasmid

**Timing: 3-5 days**

**Before Starting:** Prepare LB-agar plates containing ampicillin (100 µg/mL)
Order pX330-U6-Chimeric_BB-CBh-hSpCas9 plasmid from Addgene (#42230). This plasmid allows for dual expression of the *Streptococcus pyogenes* Cas9 and the sgRNA upon transfection of human cell lines.

First, we will describe the cloning of a gRNA into the pX330-U6-Chimeric_BB-CBh-hSpCas9 plasmid from Addgene (#42230). This plasmid allows for the simultaneous expression of the Cas9 and the chosen guide. The single-stranded DNA oligos that were designed and ordered as described in 1a will be annealed and cloned into the pX330 plasmid linearized with the restriction enzyme BpiI.

1. **Anneal guide oligos**
   a. Resuspend the oligos in nuclease-free water to a concentration of 1 mg/ml.
   b. For annealing, mix:
      4 µL sense oligo
      4 µL anti-sense oligo
      42 µL annealing buffer
   c. Using a PCR thermocycler, heat the oligo mixture to 95 °C for 2 min. and decrease the temperature by 1 °C per min. to 25 °C.

**Optional pause point:** Annealed oligos can be stored at -20 °C for future use.

2. **Purification and enzymatic digestion of pX330**
   a. Transform Top10 *E. coli* with the pX330 plasmid and plate on a LB-Ampicillin agar plate.
   b. The following day pick a colony from the plate and inoculate 5 mL of LB medium containing Ampicillin (100 mg/mL), incubate overnight (ON) at 37 °C (shaking).
   c. The following day proceed with plasmid extraction using the Wizard Plus SV Minipreps DNA Purification System.

**Note:** Purification of the pX330 plasmid can also be done in Maxiprep format for multiple uses.

   d. Digest 2 µg of the pX330 plasmid with 2 µL FastDigest BpiI in the provided digestion buffer (1X final concentration) in a total volume of 50 µL TE buffer for 60 min. at 37 °C.

**Note:** As a control, collect 1 µl of the sample before addition of the enzyme. Load 1 µl of digestion mix (before enzyme addition and after 60 min. on a 1% agarose gel. If supercoiled/undigested plasmid is visible on the gel in the 60 min. digestion condition, extend incubation time or troubleshoot your reagents.

   e. Load the entire digestion mixture on a 1% agarose gel for gel purification.

**Note:** Use a comb with larger wells when casting the gel, allowing electrophoresis of the whole digestion reaction (48 µl).

   f. Excise the DNA fragment corresponding to ~8.5 kb from the gel using a scalpel and extract the DNA from the gel using a QIAquick Gel Extraction Kit. Elute the purified plasmid in 40 µL of provided elution buffer.

**Critical: Do not dephosphorylate the digested plasmid when using non-phosphorylated oligos.**
**Alternative:** Phosphorylated annealed oligos can be cloned into a dephosphorylated, linearized pX330 plasmid.

**Optional pause point:** The digested and purified plasmid can be stored at -20 °C for future use.

3. **Ligation of annealed oligos into the linearized pX330 plasmid**
    a. Set up the ligation reaction as shown in Table 1 at room temperature (RT) for 2 h. or 4 °C ON.

**Table 1: Ligation of linearized pX330 with the annealed gRNA oligos**

<table>
<tr><th></th><th>Linearized pX330 (from Step 2)</th><th>Annealed oligos (from Step 1)</th><th>T4 DNA ligase (400,000 units/ml)</th><th>10X Ligase buffer</th><th>TE buffer</th></tr>
<tr><td>Control 1</td><td rowspan="3">1 µL (~50 ng)</td><td>-</td><td>-</td><td rowspan="3">2 µL</td><td rowspan="3">Fill up to 20 µL</td></tr>
<tr><td>Control 2</td><td>-</td><td>1 µL</td></tr>
<tr><td>Ligation</td><td>2 µL (~320 ng)</td><td>1 µL</td></tr>
</table>

**Note:** Control 1 tests the amount of undigested plasmid, control 2 will reveal the amount of self-ligating plasmid.

4. **Transformation of the ligation mixture into competent TOP10 *E. coli***
    a. Remove a 50 µL aliquot of Top10 *E. coli* from the -80 ºC freezer and thaw on ice for ~10 min.
    b. Add 5 µL of the ligation mixtures (from Step 3) to an aliquot of Top10 *E. coli.* Gently flick the tube a couple of times to mix the plasmid and the competent bacteria. Then place the tube back on ice for ~15 min.
    c. Transform *E. coli* by heat-shocking the DNA-bacteria mixture at 37 °C for 3 min.
    d. Add 500 µL LB medium (without antibiotic) to each tube, incubate for 15 min. at 37 °C (shaking).
    e. Plate bacteria on LB agar plates containing 100 µg/mL Ampicillin, incubate LB-agar plates at 37 °C ON.

5. **Screening of clones by Sanger sequencing to confirm ligation of the gRNA**
    a. Successful ligation is indicated by a few or no colonies growing on the control plates and many colonies on the ligation plate.
    b. Pick 2 colonies and inoculate 2 x 5 mL LB medium containing ampicillin (100 mg/mL) and incubate ON at 37 °C (shaking).
    c. The following day extract plasmid DNA using a Wizard Plus SV Minipreps DNA Purification System, according to the manufacturer's protocol, and measure the DNA concentration.
    d. Send 2 clones for Sanger sequencing using a U6 se primer (5' CAAGGCTGTTAGAGAGATAATTGGA 3'). This primer binds the U6 promoter region upstream of the gRNA cloning site allowing to confirm insertion of the gRNA sequence into pX330.

## 1b. Generation of HDR donor plasmids for N-terminal tagging

**Timing: 3-5 days**

Here, we will describe the single cloning step required to generate HDR donor plasmids for an N-terminally-tagged target. As described in Preparation Steps 12-13, you should have synthesized an HDR donor plasmid containing the left homology arm (LHA), followed by the NheI and BamHI restriction sites, a glycine-serin linker (L), and the right homology arm (RHA) [LHA-NheI-NNNNN-BamHI-L-RHA; NNNNN are random bases to enable efficient restriction digest]. The desired Tag and antibiotic resistance cassette combination can be simply copied and pasted from the N-terminal FAB-CRISPR plasmid collection (Addgene ID: 229676-8 and 230026; Figure 11) as a NheI-BamHI insert.

A

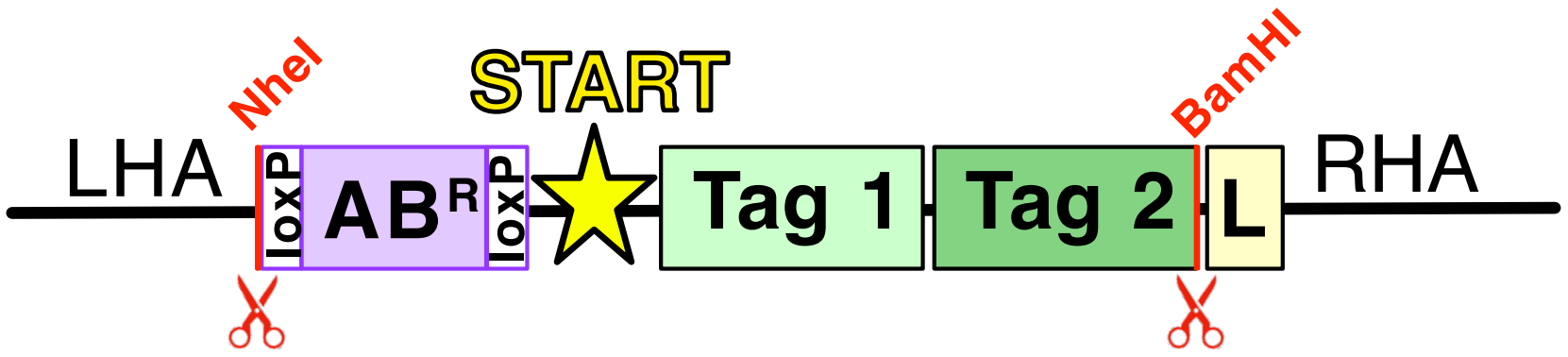

| Addgene ID | Restriction site | Antibiotic resistance ($AB^R$) cassette | Epitope tag (WB, IF) | Tag | Restriction site | Size |
|---|---|---|---|---|---|---|
| 229676 | NheI | loxP-G418-loxP | 3xALFA | Halo | BamHI | 2568 bp |
| 229677 | | loxP-Puromycin-loxP | 3xV5 | SNAP | | 2203 bp |
| 229678 | | loxP-Hygromycin-loxP | HA | mStayGold | | 2407 bp |

B

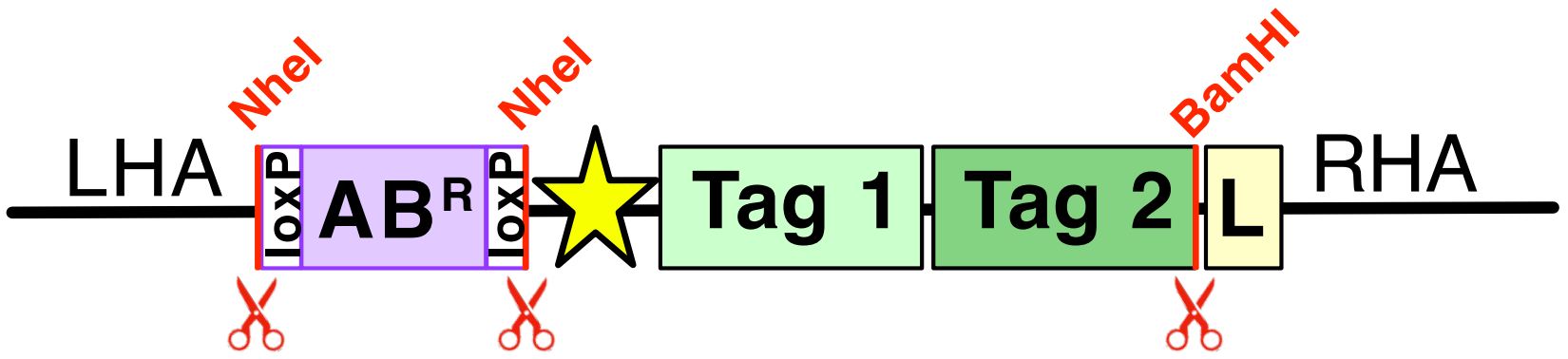

| Addgene ID | Restriction site | Antibiotic resistance ($AB^R$) cassette | Restriction site | Epitope tag (WB, IF) | Tag | Restriction site |
|---|---|---|---|---|---|---|
| 230026 | NheI | loxP-G418-loxP | NheI | V5 | Turbo-ID | BamHI |

**Size:** 1518 bp | 1014 bp

**Figure 11: Overview of the available N-terminal FAB-CRISPR inserts.** (**A-B**) For N-terminal fusions, the resistance cassette sequence is flanked by LoxP sites, allowing excision via EGFP-Cre recombinase. The FAB-CRISPR insert LoxP-G418-LoxP-V5-TurboID (B) has an additional NheI site necessitating purification and ligation of two fragments. Left homology arm (LHA), right homology arm (RHA), $AB^R$ (antibiotic resistance cassette), monomeric StayGold (mStayGold), L (glycine-serine linker). Not drawn to scale.

**Before Starting:** Prepare LB-agar plates containing ampicillin (100 µg/mL)
Order plasmids from Addgene and make DNA Minipreps.

**6. Linearization of the synthesized HDR donor plasmid**

a. Digest 2 µg of your HDR donor plasmid with 2 µL each of NheI-HF and BamHI-HF in CutSmart buffer (1X final concentration) in a total volume of 50 µl TE buffer for 60 min at 37 °C.

**Note:** To assess digestion efficiency, collect 1 µL of the sample before and after digestion. Load the samples on a 1% agarose gel. If supercoiled/undigested plasmid is visible on the gel in the digested sample, extend incubation time or troubleshoot your reagents.

b. Add 1 µL Quick CIP (5000 units/mL) to the reaction and incubate at 37 °C for 30 min.

**Note**: Quick CIP removes phosphate groups from the ends of the linearized plasmid, preventing re-ligation.

c. Clean up the digestion mix to remove the enzymes using the QIAquick PCR Purification Kit. Elute in 40 µL of the provided elution buffer.

**Optional pause point:** The digested and purified plasmid can be stored at -20 °C for future use.

**7. Generation of the N-terminal FAB-CRISPR DNA fragment**

a. Amplify the FAB-CRISPR plasmid of choice as described in Step 4a-f (see Figure 10 for an overview).

b. Digest 4 µg of your FAB-CRISPR plasmid of choice with 2 µl each of NheI-HF and BamHI-HF in CutSmart buffer (1X final concentration) in a total volume of 50 µL TE buffer for 60 min. at 37 °C.

**Note:** As a control, collect 1 µl of the sample before and after digestion. Load the samples on a 1% agarose gel. Check whether a fragment of the correct size (see Figure 11, the Rab11 backbone has a size of 3671 bp) is excised. Troubleshooting 1

c. Load the whole reaction on a 1% agarose gel for subsequent gel purification.

**Note:** Use a comb with larger wells when casting an agarose gel, allowing electrophoresis of the whole digestion reaction (48 µl).

d. Excise the desired DNA fragment (see Figure 11 for expected fragment sizes) from the gel using a scalpel and extract the DNA from the gel using the QIAquick Gel Extraction Kit and elute in 40 µl of the provided elution buffer.

**Note:** When using the FAB-CRISPR insert LoxP-G418-LoxP-V5-TurboID excise both DNA fragments (corresponding to the G418 $AB^R$ cassette and the Tag; see Figure 11B).

**Optional pause point:** The digested and purified fragments can be stored at -20 °C for future use.

**8. Ligation of the N-terminal FAB-CRISPR insert into the HDR donor plasmid**

a. Set up a ligation reaction as shown in Table 2. Ligations can be carried out at RT for 2 h. or 4 °C ON.

**Table 2: Ligation of FAB-CRISPR fragments into a linearized HDR donor plasmid**

| | Linearized HDR (from Step 6) | FAB-CRISPR Fragment (from Step 7) | T4 DNA ligase (400,000 units/ml) | 10X Ligase buffer | TE buffer |
|---|---|---|---|---|---|
| Control 1 | 1 µL (~50 ng) | - | - | 2 µL | Fill up to 20 µL |
| Control 2 | | - | 1 µL | | |
| Ligation | | 2 µL (~200 ng) | 1 µL | | |

**Note:** For the ligation of the two fragments resulting from digestion of the LoxP-G418-LoxP-V5-TurboID cassette (see Figure 11B), we recommend using 25 ng of linearized HDR plasmid and 1 µg DNA for each fragment.

b. Continue with transformation and plasmid amplification, as described in Steps 4a-f.

**Note:** Plate your transformed cells onto an antibiotic-containing plate that corresponds to the resistance cassette in your chosen backbone.

## 9. Confirming successful cloning via restriction digest and Sanger sequencing

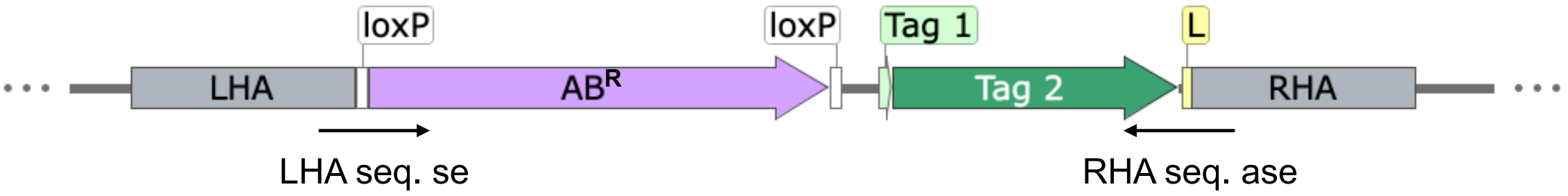

**Figure 12: Schematic of a target HDR donor plasmid for N-terminal tagging and suggested sequencing (seq) primers.** Left homology arm (LHA), right homology arm (RHA), AB$^R$ (antibiotic resistance cassette), se (sense), ase (anti-sense), L (glycine-serine linker).

a. Successful ligation is indicated by a few or no colonies growing on the control plates and many colonies on the ligation plate.
b. Pick 3 colonies and inoculate 2 x 5 mL LB medium containing ampicillin (100 mg/mL) and incubate ON at 37 °C (shaking).
c. The following day extract plasmid DNA using a Wizard Plus SV Minipreps DNA Purification System, according to the manufacturer's protocol, and measure the DNA concentration.
d. Digest 500 ng of each DNA Miniprep (from Step 8b) with 0.1 µL each of BamHI-HF and NheI-HF in a total volume of 10 µL of TE buffer
e. Load digests of DNA Minipreps from different clones on a 1% agarose gel.
f. Select the clones which display the correct digestion pattern (see Step 7b) for Sanger sequencing with a se and an ase primer that anneal with the LHA and RHA sequence, respectively. Design the primer to align ~80 bp upstream (se) and downstream (ase) from the target sequence for optimal sequencing results (Figure 12).

**Alternative:** Full-length sequencing for plasmid verification is now performed by many sequencing providers.

## 1c. Generation of HDR donor plasmids for C-terminal tagging

**Timing: 3-5 days**

Here, we will describe the single cloning step required to generate HDR donor plasmids for a C-terminally-tagged target. As described in Preparation Steps 12-13, you should have synthesized an HDR donor plasmid containing: the left homology arm (LHA), followed by a glycine-serine linker (L), a BamHI restriction site, an EcoRI restriction site and the right homology arm (RHA) [LHA-L-BamHI-NNNNN-EcoRI-RHA; NNNNN are random bases to enable restriction digests]. The desired Tag and antibiotic resistance cassette combination can be copied and pasted from the C-terminal FAB-CRISPR plasmids collection (ID: 229679-81, 227734; Figure 13). The desired insert is simply excised as a BamHI-EcoRI fragment for insertion into your newly synthesized HDR donor plasmid.

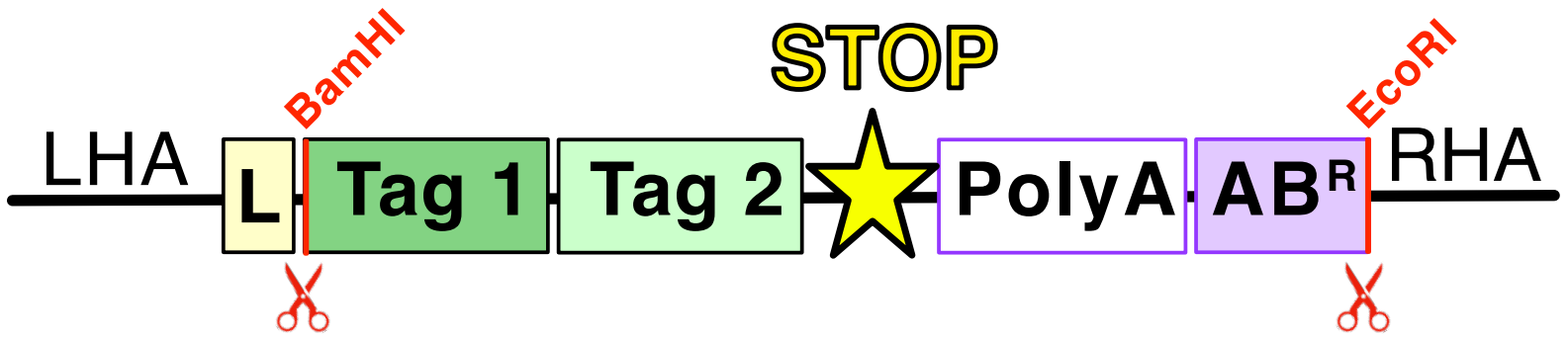

| Addgene ID | Restriction site | Tag | Epitope tag (WB, IF) | Antibiotic resistance ($AB^R$) cassette | Restriction site | Size |
|---|---|---|---|---|---|---|
| 227734 | BamHI | Turbo-ID | V5 | PolyA-G418 | EcoRI | 2722 bp |
| 229681 | | Halo | ALFA | | | 2635 bp |
| 229680 | | SNAP | V5 | PolyA-Puromycin | | 2254 bp |
| 229679 | | mStayGold | | PolyA-Hygromycin | | 2578 bp |

**Figure 13: Overview of the available C-terminal FAB-CRISPR inserts.** The presence of an exogenous polyA sequence downstream of the Tag eliminates the need to excise the resistance cassette while allowing expression of the endogenous target. Left homology arm (LHA), right homology arm (RHA), $AB^R$ (antibiotic resistance cassette), monomeric StayGold (mStayGold), L (glycine-serine linker). Not drawn to scale.

**10. Linearization of the synthesized HDR donor plasmid, generation of the FAB-CRISPR inserts and ligation**

This can be done as previously described in Steps 6-8 by simply using BamHI-HF and EcoRI-HF as restriction enzymes. Sizes of the expected fragments are listed in Figure 13, the AP1μA backbone is 4379 bp.

## 11. Confirm successful cloning via restriction digest and Sanger sequencing

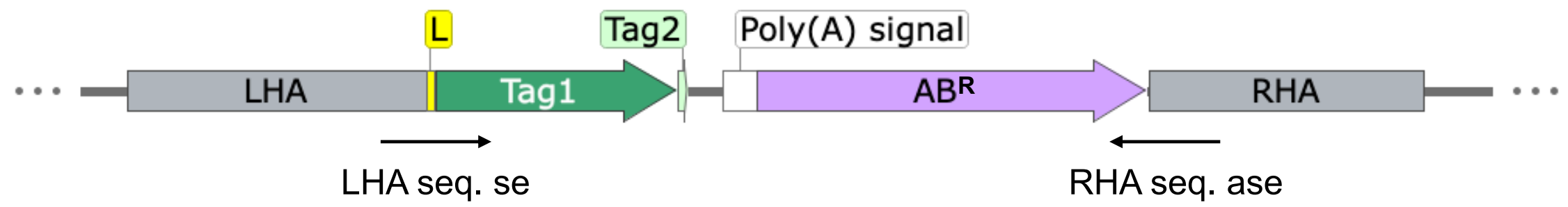

**Figure 14: Schematic of a target HDR donor plasmid for C-terminal tagging and suggested sequencing (seq) primers.** Left homology arm (LHA), right homology arm (RHA), $AB^R$ (antibiotic resistance cassette), se (sense), ase (anti-sense), L (glycine-serine linker).

a. Prepare and digest your Minipreps as described in 9a-e but using BamHI-HF and EcoRI-HF
b. Select the clones which display the correct digestion pattern (see Figure 13).
c. Select the Minipreps showing the correct digestion pattern for Sanger sequencing with a se and an ase primer that align with the LHA and RHA sequence, respectively (see Figure 14).

**Alternative:** Full-length sequencing for plasmid verification is now performed by many sequencing providers.

## 2. Transfection of editing reagents into HeLa cells

**Timing: 3 days**

For editing a gene of interest, the generated HDR donor plasmid (containing the Tag of interest and the antibiotic resistance cassette) and the pX330 plasmid (encoding for the Cas9 and the gRNA) are co-transfected into HeLa cells. HeLa cells are maintained in growth medium at 37 °C with 5% $CO_2$. A timeline of cell culture manipulations, from transfection (Step 2) to selection of positively edited cells (Major Steps 3a and 3b) is shown in Figure 15.

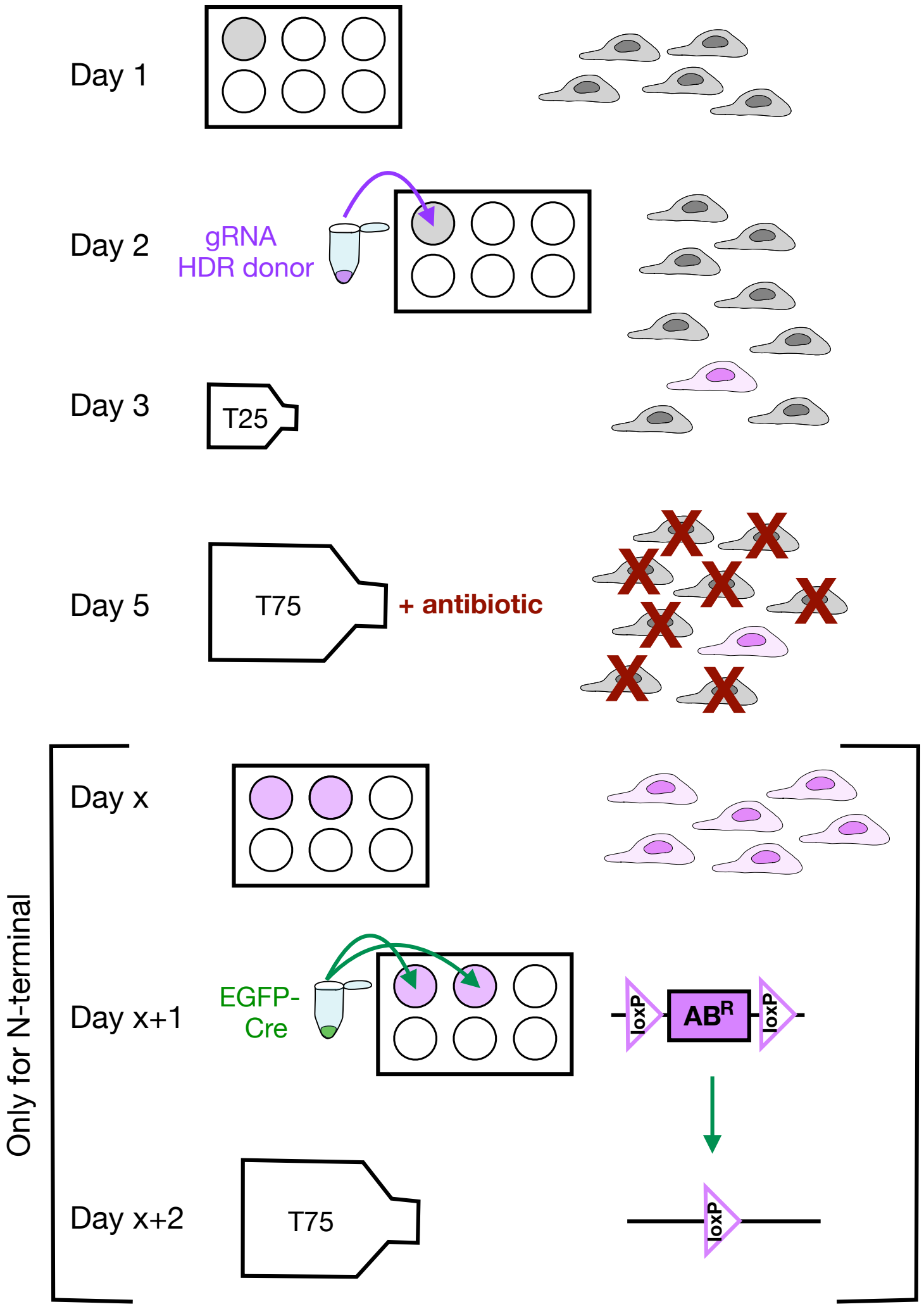

**Figure 15: Timeline for the generation of a FAB-CRISPR gene-edited cell line.** Positively edited KI cells (shown in purple) integrate an antibiotic resistance cassette into their genome, in addition to the desired Tag. KI cells can be enriched over WT cells (shown in grey) through antibiotic selection (red). When knocking in a Tag at the N-terminus of a protein of interest, the resistance cassette (flanked by LoxP sites) needs to be excised using an EGFP-Cre recombinase (green) after selection and regrowth of the cells (Day x). Not drawn to scale.

12. **Co-transfection of the FAB-CRISPR HDR donor and pX330 plasmids into HeLa cells with the transfection reagent FuGENE**
    a. Day 1: Seed cells for transfection
        i. Seed 200 000 WT HeLa cells per well into a 6-well plate. Cells should be ~60-70% confluent the following day.
    b. Day 2: Transfect cells with FAB-CRISPR constructs
        i. Prepare a transfection mixture containing: 1 µg HDR donor plasmid and 1 µg pX330 plasmid in a total volume of 100 µL OptiMEM.

ii. Vortex the mixture briefly, then add 6 µL of FuGENE (1:3 DNA:FuGENE ratio).
iii. Vortex the transfection mixture containing FuGENE, spin down briefly, incubate for 5 min. at RT.
iv. Wash the cells 3X with 2 mL PBS.
v. Add 500 µl OptiMEM per well and then apply the transfection mixture dropwise.

**Critical: Gently swirl the 6-well plate to ensure the transfection mixture is evenly distributed.**

vi. After ~5 h. exchange the medium to 2 mL growth medium, incubate ON at 37 °C with 5% $CO_2$.

c. Day 3: Transfer cells to a T25 flask
  i. Wash each well with 2 mL PBS.
  ii. Detach cells by adding 500 µL trypsin, incubate for 5 min. at 37 °C with 5% $CO_2$.
  iii. Add 1 mL of growth medium to the well, gently pipette up and down to resuspend the cells, and transfer them into a T25 flask in a total volume of 5 mL growth medium.

**Note:** For troubleshooting or optimization of the transfection reaction, follow the manufacturer's instruction.

**Note:** When creating a double-KI, simultaneously transfect pX330 and HDR donor plasmids (*e.g.,* 2 HDR plasmids containing resistance cassettes conferring resistance to both Puromycin and G418) to edit both loci (0.5 µg of each plasmid) with 6 µL FuGENE. See the alternative section (page 15) for more details.

## 3a. Selection of edited cells (C-terminal tagging)

**Timing: 1-2 weeks**

Homology-directed repair (HDR) is a rare event. To enrich for cells that have undergone HDR and have been successfully edited, we take advantage of the antibiotic resistance cassette present on the HDR donor plasmids as this is integrated into the genome with the Tag. When editing a protein of interest at the C-terminus, we employ a FAB-CRISPR cassette containing the desired Tag, followed by an exogenous polyA sequence and a resistance cassette (Bottanelli et al., 2017; see Figure 13). Successfully edited cells will express the gene conferring antibiotic resistance and can be enriched by antibiotic selection.

**Critical: Before starting the antibiotic selection, determine the minimal antibiotic concentration that is sufficient to kill unedited HeLa cells. We have found a concentration of 1 mg/mL of G418, 1 µg/mL of Puromycin and 0.4 mg/mL of Hygromycin to be optimal for selection.**

**13. Enrichment of gene-edited cells via antibiotic selection**

a. Begin antibiotic treatment three days after transfection; this gives the cells enough time to express the gRNA and Cas9 nuclease and edit the genome.
  i. Ensure cells are less than 50% confluent when first adding the antibiotic to ensure optimal selection conditions.
  ii. Replace medium with growth medium containing the appropriate antibiotic (*e.g.*, G418, Puromycin, Hygromycin).

iii. Maintain antibiotic pressure until only a few percent of the cells survive. For this, exchange the medium containing the antibiotic every 2-3 days. The antibiotic selection should be complete after one week (for G418) or 2-3 days (for Puromycin and Hygromycin). Troubleshooting 2-3

b. After antibiotic treatment, most cells should have died. The surviving cells will divide to form islands of positively-edited cells. Once these islands form, detach the cells and evenly redistribute cells in the same flask.
   i. Wash the T75 flask with 10 mL PBS.
   ii. Detach the cells by adding 2 mL trypsin, incubate for 5 min. at 37 °C with 5% $CO_2$.
   iii. Tap the flask gently to detach cells.
   iv. Add 10 mL growth medium and gently pipette up and down to break up cell clumps.

**Optional pause point**: After cells have recovered from antibiotic treatment, they can be frozen down with growth medium containing 10% DMSO for long-term storage.

## 3b. Selection of edited cells (N-terminal tagging)

**Timing: 2-3 weeks**

For N-terminal tagging, an excisable resistance cassette, flanked by LoxP sites, is edited into the genome before the desired Tag (Xi et al., 2015). The resistance cassette must be excised for the edited protein to be expressed. After cells recover from antibiotic selection, they are transfected with a plasmid coding for EGFP-Cre (Addgene plasmid #11923; Le et al., 1999) to excise the resistance cassette.

For enrichment of gene-edited cells using antibiotic treatment refer to Step 13 of the previous Major Step, “Selection of edited cells (C-terminal tagging)”.

**14. Excision of the resistance cassette with EGFP-Cre recombinase**

a. Day 1: Seed antibiotic-selected cells for transfection
   i. Seed 200 000 cells per well into a 6-well plate. Cells should be ~60-70% confluent the following day. We recommend seeding multiple wells which can be combined after transfection to obtain a confluent flask of cells more rapidly.

**Note:** Keep some antibiotic-selected cells as a backup in case the EGFP-Cre transfection needs to be repeated.

b. Day 2: Transfect cells with EGFP-Cre
   i. Prepare a transfection mixture containing 2 µg EGFP-Cre in a total volume of 100 µL OptiMEM.
   ii. Vortex the mixture briefly and add 6 µL FuGENE (1:3 DNA:FuGENE ratio).
   iii. Vortex the transfection mixture containing FuGENE and spin it down briefly, incubate for 5 min. at RT.
   iv. Wash the cells 3X with 2 mL PBS.
   v. Add 500 µL OptiMEM per well, then apply the transfection mixture dropwise.

**Critical: Gently swirl the 6-well plate to ensure the transfection mixture is evenly distributed.**

vi. After ~5 h. exchange to 2 mL of growth medium, incubate ON at 37 °C with 5% $CO_2$.

c. Day 3: Transfer cells into a T75 flask

i. Wash each well with 2 mL PBS.

ii. Detach the cells by adding 500 µL trypsin, incubate for 5 min. at 37 °C with 5% $CO_2$.

iii. Add 1 mL of growth medium in each well and resuspend the cells by gently pipetting up and down.

iv. Combine cells into a T75 flask in a total volume of 10 ml growth medium.

**Note:** To check the transfection efficiency of EGFP-Cre recombinase, examine the cells the next day using a cell culture LED-based fluorescence microscope equipped with a filter to detect green fluorescence; EGFP-Cre transfected cells will show green fluorescence in the nucleus.

## 4. Verification of successfully edited cells

**Timing: variable**

Gene-edited cells enriched via antibiotic selection can be validated through Western blotting, Sanger sequencing of the targeted genomic locus, and imaging. When clonal selection is not required, downstream experiments can be carried out with the mixed (homozygous and heterozygous) cell population. Additionally, it is important to validate the localization and functionality of the endogenously tagged protein. For the Turbo-ID KIs it is also important to validate whether the fusion protein can biotinylate proximal interactors (for more details on how to do this see (Cho et al., 2020)).

**15. Verification of endogenous tagging via Western blotting**

**Timing: 2 days**

a. Day 1: Seed cells for harvesting

i. Seed 200 000 gene-edited and 200 000 WT HeLa (negative control) cells in a 6-well plate.

b. Day 2: Sample preparation

i. The wells should be ~80% confluent on the day of harvesting.

ii. Wash the wells twice with PBS.

iii. Add 400 µL 2X Laemmli buffer containing 5% BME per well.

iv. Scrape the wells with a cell scraper and transfer the cells into a 1.5 mL tube.

v. Boil the sample for 10 min. at 95 °C.

**Optional pause point**: The boiled sample can be frozen down (-20 °C) for later use.

c. SDS Page and Western blotting

i. Vortex and load 25 µL of the sample per well in a pre-cast protein gel, alongside a molecular weight marker.

**Note:** If the sample was frozen, boil and vortex before loading.

ii. Run the SDS gel at 100-150 V until the marker has run sufficiently depending on the molecular weight of the protein of interest.

iii. Transfer the proteins to a nitrocellulose membrane by wet blotting (1 h. at 350 mA).

**Note:** To check the quality of the transfer, the membrane can be stained with Ponceau.

iv. Block the membrane with blocking buffer for 1 h. at RT (shaking).

v. Wash the membrane for 15 min. with PBST (shaking).
vi. Wash the membrane 3X for 5 min. with PBS (shaking).
vii. Add the primary antibody for 1 h. at RT or ON at 4 °C (shaking).

**Note:** Dilute the desired antibody (*e.g.*, anti-V5/ALFA/HA or anti-SNAP/Halo) antibodies 1:2000 in PBS supplemented with 0.02% sodium azide and 1% BSA.

viii. Wash the membrane for 15 min. with PBST (shaking).
ix. Wash the membrane 3X for 5 min. with PBST (shaking).
x. Add an anti-mouse or anti-rabbit HRP-conjugated secondary antibody (depending on the primary antibody used) diluted 1:5000 in blocking buffer for 45 min. at RT (shaking).
xi. Wash the membrane for 15 min. with PBST (shaking).
xii. Wash the membrane 3X for 5 min. with PBS (shaking).
xiii. Develop the blot using the SuperSignal West Pico Plus chemiluminescent substrate. Troubleshooting 4-5, Troubleshooting 8

**Note:** SuperSignal West Pico Plus is a high-sensitivity substrate which is ideal for detecting low abundance endogenously tagged proteins.

**16. In-gel fluorescent detection of endogenously tagged proteins**

**Timing: 2 days**

a. Day 1: Seed cells for harvesting
   i. Seed 60 000 cells in a 24-well plate in 500 µL growth medium.
b. Day 2: Labelling and in-gel fluorescent detection
   i. Label Halo/SNAP tagged protein with 100 nM Janelia Fluor JFX650 HaloTag or SNAP-Cell 647-SiR ligands in 500 µl growth medium, incubate for 30 min. at 37°C with 5 % $CO_2$.
   ii. Wash 3X with PBS, and then exchange to 500 µl growth medium, incubate at 37°C for 10 min. allowing any unbound dye to leach out of the cells into the medium.
   iii. Wash once with PBS and add 60 µl of 2X Laemmli buffer containing 5% BME.
   iv. Using a pipette, swirl the buffer around the well and transfer lysate to a 1.5 ml tube.
   v. Boil the sample at 95°C for 5 min. and load 25 µl onto a polyacrylamide gel alongside a molecular weight marker.
   vi. Run the gel at 200 V for ~45 min. until the Laemmli buffer runs out of the bottom of the cassette.
   vii. Using the Cy5.5 filter on a BioRad ChemiDoc, image fluorescence (1-60 s. depending on the expression of your tagged protein). Troubleshooting 5, Troubleshooting 8

**17. Sanger sequencing of the targeted genomic locus**

**Timing: 2 days**

a. Day 1: Seed cells for harvesting
   i. Seed 10 000 gene-edited and 10 000 WT HeLa (negative control) cells in a 6-well plate.
b. Day 2: Extract DNA
   i. Wash the wells 3X with PBS.

ii. Add 500 µL QuickExtract DNA Extraction Solution and transfer the cells using a cell scraper into a 1.5 mL tube.
iii. Vortex the 1.5 mL tube for 15 s, incubate for 6 min. at 65 °C.
iv. Vortex the 1.5 mL tube for 15 s, incubate for 2 min. at 98 °C.

**Note:** For details see www.lucigen.com/docs/manuals/MA150E-QuickExtract-DNA-Solution.pdf.

c. Genomic PCR
   i. Set up a genomic PCR with the extracted genomic DNA and primers binding outside the HAs (see Table 3 for a detailed reaction and Table 4 for cycling conditions).

**Table 3: PCR reaction mix**

| Reagent | Amount |
|---|---|
| 5X Phusion Reaction buffer | 5 µL |
| dNTPs | 0.5 µL |
| Sense primer (10 mM) | 1.25 µL |
| Antisense primer (10 mM) | 1.25 µL |
| Template | 25 ng |
| Phusion polymerase | 0.25 µL |
| Millipore water | Up to 25 µL |

**Table 4: PCR cycling conditions**

| Steps | Temperature | Time | Cycles |
|---|---|---|---|
| Initial Denaturation | 98 °C | 5 min. | 1 |
| Denaturation | 98 °C | 30 s. | 45 |
| Annealing | x | 1 min. | |
| Extension | 72 °C | 2 min.* | |
| Final extension | 72 °C | 5 min. | 1 |
| Hold | 4 °C | ∞ | |

x= Use recommended temperature of a Tm calculator (*e.g.*, https://tmcalculator.neb.com).
* = Adjust to fragment size and according to polymerase manufacturer instructions.

d. Gel purify the PCR fragment with a gel extraction kit and send the DNA for Sanger sequencing (see Figure 15). Troubleshooting 8

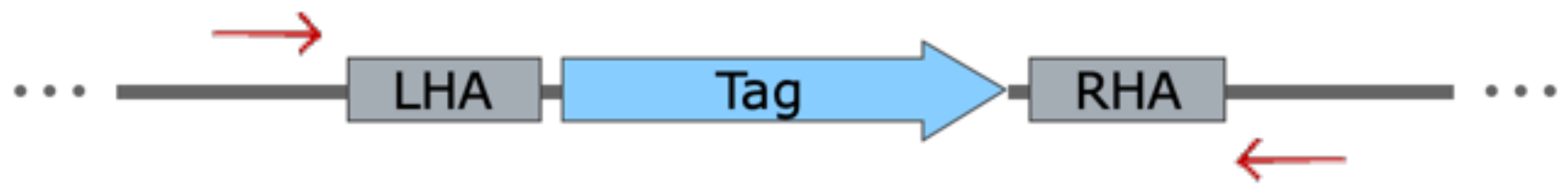

**Figure 15: Schematic showing how to design PCR primers (red arrows) to amplify the targeted genomic locus.** Left homology arm (LHA), right homology arm (RHA). Not drawn to scale.

18. **Verification of the expression and localization of mStayGold-, Halo-, SNAP-tagged proteins with live-cell imaging**

**Timing: 2 days**

Gene-edited cells can be visualized via live-cell imaging using fluorescence (mStayGold) or by labelling the self-labelling enzymes SNAP and Halo with cell permeable substrates.

a. Day 1: Seed cells for imaging
    i. Seed 160 000 gene-edited cells on a glass bottom 35 mm dish in a total volume of 2 mL growth medium. For optimal imaging, ensure cells are ~70% confluent the following day.
b. Day 2: Stain Halo/SNAP tagged cells
    i. Add 1 µM ligand (Janelia Fluor JFX650 HaloTag or SNAP-Cell 647-SiR) in a total volume of 120 µL growth medium in the 14 mm micro-well at the center of the dish, incubate for 1 h. at 37 °C with 5% $CO_2$.
    ii. Wash 3X with 2 mL growth medium
    iii. Incubate for 60 min. at 37 °C with 5% $CO_2$ to allow any unbound dye to leach out of the cells into the medium.

**Note:** SNAP and Halo substrates are available in a palette of colors from New England Biolabs (SNAP) and Promega (Halo). For alternative SNAP and Halo labelling strategies see (Broadbent et al., 2023; Heyza et al., 2023a; Stockhammer et al., 2024a; Wong-Dilworth et al., 2023).

**Note:** mStayGold is a fluorescent protein and does not require staining.

    iv. Shortly before imaging, wash cells with 2 mL PBS and add 2 mL pre-warmed live-cell imaging solution.
    v. Image cells on a light microscope using settings to detect fluorophores in the far-red range (for Janelia Fluor JFX650 HaloTag or SNAP-Cell 647-SiR ligands; 640-650 nm laser) or green range (for mStayGold; 488 nm laser) to assess the localization of the protein of interest. We recommend using a light microscope with laser sources as high laser intensities may be required to visualize low abundance proteins. Troubleshooting 5-8

**Alternative:** A primary antibody against the inserted Tag (for recommended anti-V5/ALFA/HA/Halo/SNAP antibodies see Key Resource Table) in combination with a secondary antibody against the primary antibody carrying a fluorophore can be used for immunofluorescence.

**Alternative:** For the Turbo-ID KIs use a primary anti-V5 antibody in combination with a secondary antibody against the primary antibody carrying a fluorophore for immunofluorescence.

## 5. Further selection of edited cells

**Timing: variable**

If the editing efficiency is insufficient or homozygous clones are required, cells edited with SNAP, Halo or mStayGold can be further enriched through fluorescence activated cell sorting (FACS). Following FACS enrichment, cells can be validated via Western blotting (see Step 15), in-gel fluorescent detection of the tagged proteins (Step 16) and imaging (Step 18). Clones derived from single cells can be genotyped to assess zygosity (see Step 20).

**19. Further enrichment for edited cell with FACS**

a. Seed cells for FACS
   i. Seed edited cells in a T75 flask; a confluent flask is required on the day of sorting.
b. Cell preparation for FACS
   i. On the day of sorting, label gene-edited cells with 0.5 µM SNAP/Halo ligands in growth medium as described in Step 18b. Typically, we stain a confluent T75 flask in 5 mL of growth medium to reduce dye usage.
   ii. Sort labelled cells via FACS, gating on the 0.5-1% brightest cells to select for homozygous edited cells. For single-cell sorting, single cells can be collected in each well of a 96-well plate and grown to confluency.

**20. Genotyping of clonal cell lines derived from single cells**

a. Amplify the target locus as described in Step 16.
b. Load 5 µL of the PCR product on an agarose gel. Use WT cells as a negative control.
c. Assess the DNA patterns on the gel. Amplification of the WT locus should yield a low molecular weight band (see WT lane in Figure 16 of an example gel picture). Amplification of the edited locus should yield a higher molecular weight band because of the insertion of a large fragment (*e.g.,* Halo) via editing. Heterozygous clones will show 2 bands (one for the edited allele and one for the WT allele, lane cl.1) while homozygous clones will show a single higher molecular weight band as both alleles are edited (lane cl.2).

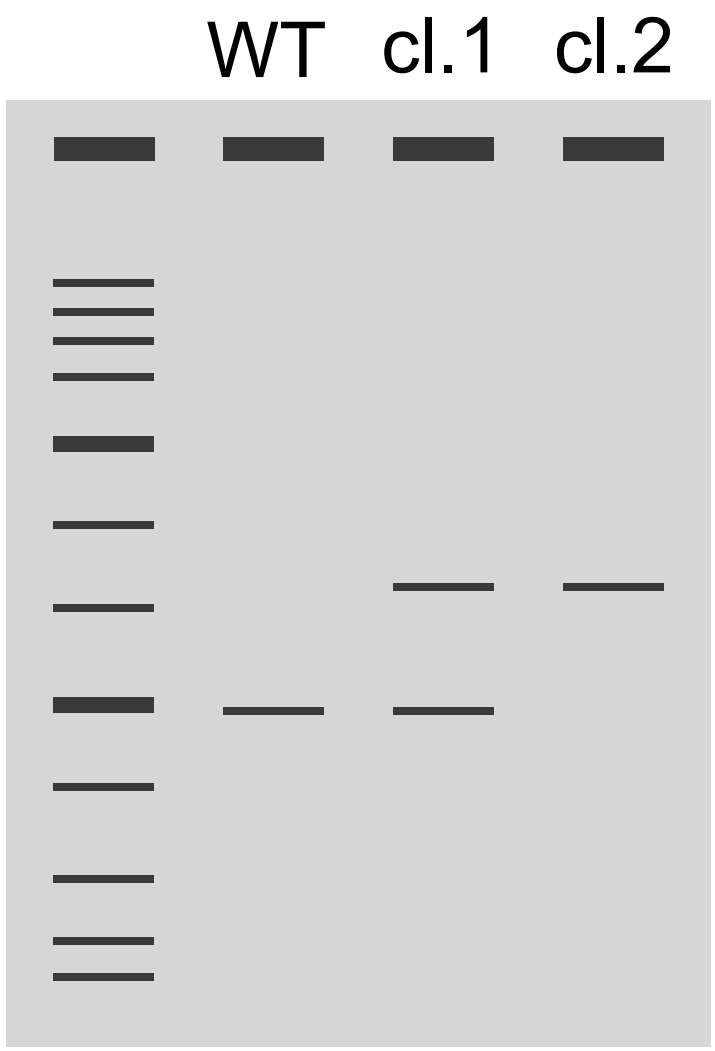

**Figure 16: An example agarose gel image for CRISPR clone genotyping.** Wild type (WT) serves as a control for the unedited protein loaded next to CRISPR-Cas9 clone 1 (cl. 1, heterozygous) and 2 (cl.2, homozygous).

**Critical: When working with clones derived from single cells, we recommend using several clones for downstream experiments to ensure any observed effects are not due to clonal selection.**

## 7. Functional verification of the endogenously tagged protein

**Timing: variable**

To ensure the Tag does not disrupt protein function, additional experiments are necessary. The required experiments depend on the biological role of the protein. For example, trafficking assays are suitable to test the functionality of proteins involved in membrane trafficking. Verification of the functionality of the tagged protein of interest must be carried out on homozygous clones where both alleles have been edited. Alternatively, a haploid cell line like HAP1 can be used, where a single editing event yields 100% tagged products (Essletzbichler et al., 2014; Wong-Dilworth et al., 2023).

# Expected outcomes

C-terminal tagging with FAB-CRISPR should yield KI efficiencies (percentage of edited cells in a population of cells) close to 100%. Some variability is expected depending on the chosen targeting gRNA. For N-terminal tagging the final KI efficiency will depend on the transfection efficiency of the EGFP-Cre recombinase plasmid used to excise the antibiotic resistance cassette. However, efficiencies as high as 80% can be achieved when both transfection steps are optimally performed. We recommend reading the troubleshooting section for tips on how to improve KI efficiencies.

# Limitations

- Endogenous expression levels are moderate when compared to expression levels achieved via plasmid overexpression. This needs to be accounted for when performing downstream experiments (*e.g.,* for mass spectrometry experiments one may have to increase the amount of cell material used). Various tools are available for checking protein and transcript abundance (but bear in mind that mRNA levels do not always correlate with protein levels!).
    1. http://mapofthecell.biochem.mpg.de/index.html (Itzhak et al., 2016)
    2. https://www.proteinatlas.org/ (Uhlen et al., 2010)
- When using a non-fluorescent Tag, such as Turbo-ID, FACS is not possible and single clone selection will have to be carried out manually via serial dilution.
- Tagging at the N-terminus using the excisable resistance cassette strategy leads to a transient KO of the protein of interest as the locus is disrupted. Upon excision of the resistance cassette using EGFP-Cre, expression of the target protein should be restored (validate your cell lines using the suggested methods). For essential genes, this strategy will likely only yield heterozygous KIs, as homozygous KIs will not be viable.
- When trying to edit essential genes, tagging at the C-terminus is less likely to disrupt protein expression because the protein is likely expressed during the antibiotic selection. Tagging the N-terminus of essential genes can cause a loss of protein functionality. If it must be tagged N-terminally, we advise direct N-terminal insertion of the Tag devoid of the LoxP-AB$^{R}$-LoxP cassette. This strategy overcomes the temporary non-functionality of the expressed protein. A caveat of direct insertion is that the lack of the antibiotic selection marker may reduce the knock-in efficiency. In this case, FACS (Step 19) can then be used to enrich for the gene-edited cells.

- In the design provided in this protocol, the LoxP-AB$^R$-LoxP cassette is inserted upstream of the endogenous ATG start codon. To avoid changing any bases prior to the start codon, since they may contain sequences that are important for transcription and translational efficiency, we have successfully inserted the LoxP-flanked cassette downstream of the endogenous start codon (Xi et al., 2015).

# Troubleshooting

## Problem 1:

Unexpected fragment pattern observed in HDR donor plasmid digestion (Step 7).

## Potential solution:

Ensure the LHA, RHA, AB$^R$, Tags, and plasmid backbone sequences do not include restriction sites needed for cloning of the final HDR donor plasmid. This can be done by performing an *in silico* digest on the assembled target sequence with SnapGene or a molecular biology tool of choice. If a restriction site is present within an intron or coding sequence, the sequence in the HDR donor plasmid can be mutagenized. Within an intron, beware of changing bases at the exon-intron junctions. Within the coding sequence, insert a silent mutation to not change the aminoacidic sequence. If this is not feasible, desired restriction sites can be introduced by PCR amplification of the FAB-CRISPR inserts (use FAB-CRISPR plasmids as a template) with primers that encode the restriction sites, as well as a 20 bp overlap with the beginning (se primer) or end (ase primer) of the FAB-CRISPR insert.

## Problem 2:

No cells survive the antibiotic treatment (Step 13).

## Potential solutions:

- Troubleshoot the transfection method with a fluorescent marker (*e.g.*, EGFP-tagged overexpression protein).
- Check the quality of the used DNA Mini/Maxipreps.
- When tagging at the N-terminus using the excisable resistance cassette strategy, the tagged protein won't be expressed until the resistance cassette is properly excised. Thus, the lack of surviving cells may indicate that the targeted gene is essential for cell viability.
- Check the cutting efficiency of your gRNAs with a surveyor assay (Ran et al., 2013).

## Problem 3:

No cells are killed by the antibiotic treatment (Step 13).

## Potential solution:

If the cells grew too confluent during the antibiotic treatment, the selection efficiency may be reduced (we have particularly observed this when using G418). In this case, repeat from Step 12 onward and ensure that the cells are less than 50% confluent when starting the antibiotic selection. Trypsinization of the cells prior antibiotic addition improves selection. If necessary, increase the antibiotic concentration (for G418 up to 3 mg/mL) during the first few days of selection.

## Problem 4:

Replacement of endogenous with an exogenous polyA sequence (C-terminal tagging) may alter protein expressions when the amount of transcript is tightly regulated at the mRNA level (Before you begin, HDR donor plasmid design). This can result in altered expression levels (protein levels significantly higher or lower than the endogenous expression levels detected in WT cells) of C-terminally tagged proteins (Step 15).

## Potential solution:

Construct your HDR donor plasmid to include an excisable resistance cassette (LoxP-$AB^R$-LoxP) downstream of the STOP codon of the Tag. This allows the resistance cassette to be removed post-selection.

## Problem 5:

No endogenously tagged protein is detected via Western blotting, in-gel fluorescence and/or via fluorescence microscopy (Step 15, 16, 18).

## Potential solution:

- Adding a high-affinity epitope Tag along with Halo/SNAP/mStayGold (ALFA/V5/HA) may simplify detection and allow signal amplification via immunofluorescence and/or Western blotting.
- Use high sensitivity Western blot detection reagents.
- Optimize excitation and detection settings on the fluorescence microscope.

## Problem 6:

Your Halo/SNAP-tagged protein is expressed at very low endogenous levels and the background of cell-permeable Halo/SNAP substrates (Step 18) masks the real signal (*e.g.,* bright structures due to endocytosed dye).

## Potential solution:

- Test the amount of dye background by staining unedited WT cells.
- Leave excess dye to wash out for longer times (> 1h).
- Test the localization of the protein of interest tagged with mStayGold.

## Problem 7:

Tagging interferes with protein localization and function (Step 18).

## Potential solution:

- Increase the length of the linker between the target protein and the Tag for example by using a localization and affinity purification (LAP) Tag linker (Cheeseman and Desai, 2005; Wong-Dilworth et al., 2023).
- If working with single cell clones, we recommend performing the same experiments using additional clones to rule out clonal differences.

## Problem 8:

The CRISPR KI efficiency is too low (Step 15-18).

## Potential solution:

Low KI efficiency may be due to different factors:

- Antibiotic selection has not worked efficiently (see Problem 4).
- The gRNA is not cutting efficiently. For this, test other gRNAs and test their efficiency of gRNA targeting with the surveyor assay (Ran et al., 2013)
- For N-terminally tagged proteins, inefficient excision of the LoxP-AB$^R$-LoxP cassette by the EGFP-Cre recombinase lowers the percentage of cells expressing the tagged proteins. In this case, we recommend troubleshooting the transfection of the EGFP-Cre recombinase by monitoring nuclear EGFP expression.

# Resource availability

### Lead contacts

Francesca Bottanelli: francesca.bottanelli@fu-berlin.de
Jens Schmidt: schmi706@msu.edu

### Materials availability

*Plasmids generated in this study have been deposited to Addgene and all catalogue numbers are indicated in the resources table.*

### Data and code availability

*This study did not generate/analyze any datasets/code.*

### Acknowledgments

We acknowledge all the members of the Bottanelli and Schmidt laboratories for supporting the development of the tools explained in this protocol. P.A is supported by the Deutsche Forschungsgemeinschaft (DFG) – project number 278001972 – TRR 186.

# Author contributions

All authors contributed to the writing of this manuscript.

# Declaration of interests

The authors declare no competing interests.

## Annex 1: Cloning of the HDR plasmid via Gibson assembly

**Timing: 5 days**

Here we will describe step-by-step how to construct an HDR donor plasmid from double-stranded DNA fragments (dsDNA) fragments via Gibson assembly. First, left and right homology arms are either synthesized as dsDNA as described in the "Designing a homology-directed repair donor (HDR) plasmid for N- and C-terminal tagging" chapter or amplified via PCR from genomic DNA. Second, a Tag fragment is generated via PCR. Finally, the three fragments (LHA-Tag-RHA) are assembled into a vector via Gibson assembly. The generated HDR donor plasmid can then be then used for transfection and editing, as laid out from Step 2. The following steps are described in detail below.

1. Design of the left and right homology arms
2. Generation of the Tag fragment
3. Linearization of the vector
4. Assembly of the HDR template from dsDNA fragments
5. Transformation of the assembled HDR template
6. Restriction digestion for verifying correct assembly of the HDR donor plasmid

**1. Design of the left and right homology arms**

For each target gene, two dsDNA fragment encoding the right and left homology arms (HA) flanking the tag insertion site must be designed using Benchling (see the "Designing a homology-directed repair donor (HDR) plasmid for N- and C-terminal tagging" chapter). We have successfully used HA, which are of lengths 250-450 bp downstream/upstream of the START codon (N-terminal tagging) or downstream/upstream from the STOP codon (C-terminal tagging). The HA will be cloned within the HDR vector using Gibson assembly and must contain a sequence of 20-25 bp on both sides that overlap with the HDR plasmid backbone and the Tag sequence.

a. Identify the vector used for cloning the HDR donor. Ideally, the vector should be a high copy number plasmid containing a multi cloning site (MCS) for restriction digestion. In our lab, we have used the pFastBac Dual vector (ThermoFisher #10712024). Other high-copy plasmids are suitable for HDR cloning, including pUC cloning vectors.
b. The HA can be obtained as a double-stranded DNA fragment from nucleic acid suppliers. In some instances (*e.g.,* GC-rich sequences) DNA synthesis is not achievable. In this case, HA can be obtained through the amplification of genomic DNA using PCR.
c. When ordering HA as double-stranded DNA fragments, make sure to add overlapping sequences with the adjacent gene fragment/vector. When using pFastBac Dual, we linearize the plasmid using the HpaI restriction enzyme (NEB) so the LHA will have 20-25 bp overlapping with the vector plasmid on the 5' end of the HpaI cut site, and 20-25 bp overlapping with the Tag. Similarly, the RHA will have 20-25 bp on the 5' end overlapping with the Tag, and 20-25 bp on the 3'-end of the HpaI cut site overlapping with the vector.

**Note:** If the HA is PCR-amplified from genomic DNA, use SnapGene or analogous cloning software to design the amplification primers. The primers must contain 20-25 bp overlapping with the adjacent gene fragment or vector to be cloned using Gibson assembly.

d. Before proceeding with the HDR plasmid cloning, perform an *in silico* Gibson assembly using cloning design software, such as SnapGene or Benchling. For the N-terminal Tag, verify that the *in silico* assembled HDR plasmid contains a START codon immediately preceding the Tag. For the C-terminal, verify the presence of a STOP codon at the end of the Tag fragment.

**Alternative:** To overcome PCR-amplification of HA from genomic DNA when HAs are difficult to synthesize, we have had success shortening the HA by 50-200 bp (total length of homology 300-400 bp). However, this strategy could reduce the efficiency of homology-directed repair.
**Critical: If the gRNA does not span the Cas9 cleavage site, the Cas9 will cleave the HDR donor plasmid. To overcome this issue, insert a silent mutation(s) in the homologous donor sequence that will prevent the gRNA from binding.**

**2. Generation of the Tag fragment: Tag amplification and purification**

a. Obtain the Tag sequence according to your knock-in (KI) strategy. The plasmid encoding the N-terminal Halo Tag can be ordered from Addgene (#86843). The sequence of the C-terminal tag and N-terminal direct-insertion can be found as supplemental material of this protocol.
b. Using cloning software, design primers to amplify the Tag. The sense primer will contain 20 - 25 bp on the 5'-end overlapping with the LHA on the 3'- end. For the antisense primer, a 3'-end overlap is not needed if included in the 5'-end of the RHA.

**Note:** Primers with 20 - 25 bp overlap with the HA are not required if the overlapping sequences with the Tag are included when designing the HA.
**Note:** For optimal amplification, the primers should have similar $T_m$ for efficient PCR.

c. Perform the PCR and purify the product after electrophoresis on 1% agarose gel using a gel extraction kit and elute in at least 30 µL elution buffer.
d. Measure the concentration of the purified Tag PCR product.
e. Dilute the purified Tag PCR product to a concentration of ~20 ng/mL.

**3. Linearization of the vector: pFastBac Dual Digestion and Purification**

a. Digest the pFastBac Dual plasmid using HpaI restriction enzyme by assembling the reaction listed in Table 1 in a 200 µL PCR tube, incubate for 60 min. at 37 °C in a thermocycler.

**Note:** In the case of a different vector, adjust the protocol accordingly to the restriction enzyme used.

**Table 1: Digestion reaction mix**

| Component | Volume (µL) |
|---|---|
| pFastBac (1 mg/mL) | 10 |
| 10x CutSmart Buffer | 5 |
| HpaI-HF | 3 |
| $ddH_2O$ | 32 |
| Total | 50 |

**Critical: Because HpaI generates blunt DNA ends, dephosphorylation of 5'-ends of vector DNA is critical to prevent re-ligation of the backbone.**

b. Perform the dephosphorylation step immediately after HpaI digestion using Quick CIP. Add 2 µl of CIP to the enzyme digestion preparation and incubate at 37 °C for 30 min.
c. Stop the reaction at 80 °C for 2 min.
d. Purify the digested vector after electrophoresis on 1% agarose gel using a gel extraction kit and elute in at least 25 µL elution buffer. In the electrophoresis, run a small amount of undigested pFastBac vector alongside the digested vector as a control to confirm digestion.
e. Measure the concentration of the digested vector and dilute it to 50 ng/mL.

**4. Assembly of the HDR template from dsDNA fragments**

a. Dilute the HA in TE Buffer to a concentration of 10 ng/mL following the protocol provided by the supplier.
b. Set up the Gibson assembly reaction, using the Gibson Assembly® MasterMix or the NEBuilder HiFi DNA Assembly Master Mix from NEB (#E2611 or #E2621) and according to manufacturer instructions

**Note:** For an efficient Gibson assembly reaction, you want to use an excess molar ratio of each insert relative to the starting concentration of the DNA backbone. We have had much success using a 1:3 molar ratio between vector and inserts. For calculating the volume of each insert needed for the Gibson Assembly reaction, use the NEBioCalculator (https://nebiocalculator.neb.com/#!/ligation).
**Note:** As a negative control, omit all the DNA inserts and compensate the volume with $ddH_2O$.

**Table 2: Gibson assembly pipetting scheme**

| Component | Volume (µL) |
|---|---|
| Digested pFastBac (50 ng/µL) | 1.25 |
| LHA (10 ng/µL) | 1 |
| RHA (10 ng/µL) | 1 |
| Tag (20 ng/µL) | 2.5 |
| Gibson Mastermix® (2X) | 10 |
| ddH2O | 4.5 |
| Total | 20 |

c. Prepare the reaction listed in Table 2 in a 200 µL PCR tube, gently mix and incubate at 50 °C for 1h in a thermocycler, and then ramp down to 4°C.
d. Following incubation, store samples on ice or at -20 °C for subsequent transformation.

**5. Transformation of the assembled HDR template**

e. Add 5 µl of the Gibson assembly mixture to 50 µL *E. coli* TOP10, incubate for 25 min on ice.
f. Heat-shock the cells at 42 °C for 45 s.
g. Incubate on ice for 3 min.
h. Plate the cells on ampicillin-containing agar plates, incubate at 37 °C for 16–18 h.

**CRITICAL: Count the colony numbers. Usually, an efficient Gibson assembly will result in the absence of colonies in the control plate and 10-50 colonies in the assembly plate.**

**If the control and assembly plate have a similar number of colonies, repeat the Gibson assembly. Colonies in the control plate may also indicate incomplete restriction digestion of the vector plasmid. In that case, it is recommended to repeat the restriction digestion and subsequent purification.**

**6. Restriction digestion for verifying correct assembly of the HDR donor plasmid**
This step is optional and not necessary if significantly more colonies on the plus insert plate.

a. Select 2-5 colonies and inoculate 5 mL LB Broth containing ampicillin (100 mg/mL) and place in a bacterial shaker overnight at 250 rpm.
b. The following day extract plasmid from *E. coli* using a Miniprep plasmid extraction kit and measure the DNA concentration.
c. Digest the pFastBac Dual plasmid using a restriction enzyme by assembling the reaction listed in Table 3.

**Note:** When using the pFastBac, we have used BamHI enzyme, which cuts both the vector backbone and at a BamHI site in both the N- and C-Terminal Halo Tag sequence between the HA. In the case of a different vector or Tag, adjust the protocol accordingly to the restriction enzyme used.

**Table 3: Digestion reaction mix**

| Component | Volume (µL) |
|---|---|
| pFastBac (1 mg/mL) | 1 |
| 10x CutSmart Buffer | 2 |
| BamHI | 1 |
| dd$H_2O$ | 16 |
| Total | 20 |

d. Run the PCR product on a 1% agarose gel and verify that the digested product contains two discrete bands. The lack of HDR insertion (likely annealed backbone) will show a single band.
e. Design and order primers for sequencing the entire HDR DNA that you have cloned into pFastBac Dual.
f. Send the HDR plasmid for sequencing and confirm by importing Sanger sequencing results into SnapGene and aligning to your *in silico* Gibson assembled plasmid.

**ALTERNATIVE:** Instead of restriction enzyme digestion, the correct assembly of the HDR donor plasmid can be verified by colony PCR (follow your polymerase of choice provider instructions). The amplification of a fragment within the LHA-Tag-RHA will indicate the presence of the HDR sequence. As an alternative to multiple Sanger sequencing, full-length plasmid sequencing is now widely available.